\documentclass[draftclsnofoot,peerreview,english]{IEEEtran}
\usepackage[T1]{fontenc}
\usepackage[latin9]{inputenc}
\usepackage{amsmath}
\usepackage{amssymb}
\usepackage{mathrsfs}
\usepackage{bbm}
\usepackage{mathtools}
\usepackage{empheq}
\usepackage{graphicx}
\usepackage{amsthm}	
\usepackage{hyperref}
\usepackage{cite}


\usepackage{babel}

\newtheorem{thm}{Theorem}				
\newtheorem{prop}[thm]{Proposition}		
\newtheorem{lem}[thm]{Lemma}			
\newtheorem{cor}[thm]{Corollary}

\DeclareMathOperator{\pr}{\mathrm{Pr}} 		

\begin{document}

\title{The Unbounded Benefit of Encoder Cooperation for the $k$-user MAC}

\author{Parham~Noorzad,~\IEEEmembership{Student Member,~IEEE,} 
Michelle Effros,~\IEEEmembership{Fellow,~IEEE,}\\ and Michael
Langberg,~\IEEEmembership{Senior~Member,~IEEE}%
\thanks{This paper was presented in part at the 2015 IEEE International 
Symposium of Information Theory (ISIT) in Hong Kong \cite{Noorzad2} 
and the 2016 IEEE ISIT in Barcelona, Spain \cite{kUserMAC}.}%
\thanks{This material is based upon work supported by the 
National Science Foundation under Grant Numbers 15727524,
1526771, and 1321129. }%
\thanks{P. Noorzad and M. Effros are with the California
Institute of Technology, Pasadena, CA 91125 USA
(emails: parham@caltech.edu, effros@caltech.edu). }%
\thanks{M. Langberg is with the State
University of New York at Buffalo, Buffalo, NY 14260 USA
(email: mikel@buffalo.edu).}}

\maketitle

\begin{abstract}
Cooperation strategies allow communication devices to work 
together to improve network capacity. Consider a network
consisting of a $k$-user multiple access channel (MAC) and a node
that is connected to all $k$ encoders via rate-limited bidirectional 
links, referred to as the ``cooperation facilitator'' (CF). 
Define the cooperation benefit as the sum-capacity gain 
resulting from the communication between the encoders and the 
CF and the cooperation rate as the total rate the CF shares 
with the encoders. This 
work demonstrates the existence of a class of $k$-user MACs 
where the ratio of the cooperation benefit to cooperation rate
tends to infinity as the cooperation rate tends to zero. 
Examples of channels in this class
include the binary erasure MAC for $k=2$ and the 
$k$-user Gaussian MAC for any $k\geq 2$.
\end{abstract}


\section{Introduction} \label{sec:intro}

In large networks, resources may not always be distributed 
evenly across the network. There may be times where parts 
of a network are underutilized, while others
are overconstrained, leading to suboptimal performance. 
In such situations, end users are not able to use
their devices to their full capabilities.

One approach to address this problem 
allows some nodes in the network to ``cooperate,'' that is, work
together, either directly or indirectly, to achieve common goals. 
The model we next introduce is based on this idea. 

In the classical $k$-user multiple access channel (MAC) \cite{Ulrey},
there are $k$ encoders and a single decoder. Each encoder has a private
message which it transmits over $n$ channel uses to the decoder.
The decoder, once it receives $n$ output symbols, finds the messages of 
all $k$ encoders with small average probability of error. In this model, 
the encoders cannot cooperate, since each encoder only 
has access to its own message. 

\begin{figure} 
	\begin{center}
		\includegraphics[scale=0.25]{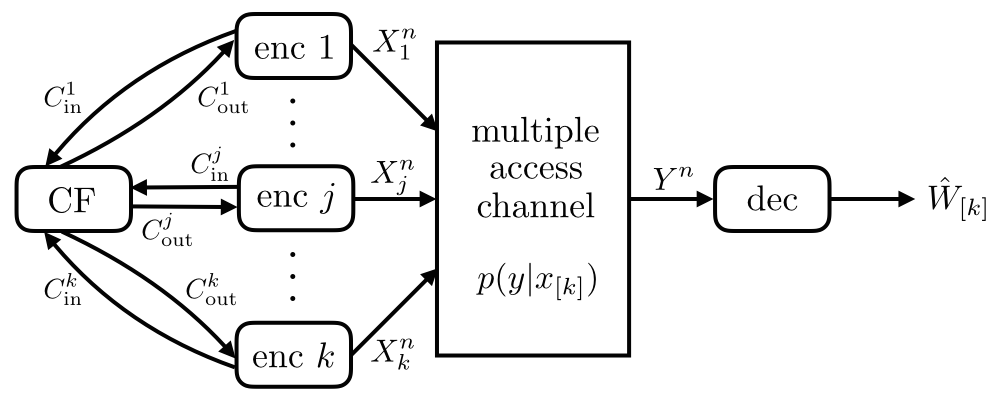}
		\caption{The network consisting of a $k$-user
		MAC and a CF. For $j\in [k]$, encoder $j$ has 
		access to message $w_j\in [2^{nR_j}]$.} \label{fig:network}
	\end{center}
\end{figure}

We now consider an alternative scenario where our $k$-user MAC is part
of a larger network. In this network, there is a node that is connected to 
all $k$ encoders and acts as a ``cooperation facilitator'' (CF). 
Specifically, for every 
$j\in [k]$,\footnote{The notation $[x]$ describes the set 
$\{1,\dots,\lfloor x\rfloor\}$ for any real number $x\geq 1$.}
there is a link of capacity $C_\mathrm{in}^j\geq 0$ going from 
encoder $j$ to the CF and a link of capacity $C_\mathrm{out}^j\geq 0$
going back. The CF helps the encoders exchange 
information before they transmit their codewords over the MAC. 
Figure \ref{fig:network} depicts a network consisting of a $k$-user MAC 
and a $(\mathbf{C}_\mathrm{in},\mathbf{C}_\mathrm{out})$-CF, where
$\mathbf{C}_\mathrm{in}=(C_\mathrm{in}^j)_{j\in [k]}$ and
$\mathbf{C}_\mathrm{out}=(C_\mathrm{out}^j)_{j\in [k]}$
denote the capacities of the CF input and output links. 
In this figure, $X^{n}_{[k]}=(X^{n}_1,\dots,X^{n}_k)$ 
is the vector of the channel inputs of the $k$ encoders, and 
$\hat{W}_{[k]}=(\hat{W}_1,\dots,\hat{W}_k)$ 
is the vector of message reproductions at the decoder. 

The communication between the CF and the encoders occurs over a number
of rounds. In the first round of cooperation, each encoder sends a rate-limited
function of its message to the CF, and the CF sends a rate-limited function
of what it receives back to each encoder. Communication between the encoders 
and the CF may continue for a finite number of rounds, with each node 
potentially using information received in prior rounds to determine 
its next transmission.  Once the communication between the CF and 
the encoders is done, each encoder uses its message and what it has 
learned through the CF to choose a codeword, which it transmits across the channel.
  
Our main result (Theorem \ref{thm:sumCapacity}) determines a set of MACs where the 
benefit of encoder cooperation through a CF grows very quickly 
with $\mathbf{C}_{\mathrm{out}}$. Specifically, we find a class of MACs 
$\mathcal{C}^*$,
where every MAC in $\mathcal{C}^*$ has the property that for any fixed 
$\mathbf{C}_\mathrm{in}\in\mathbb{R}^k_{>0}$, the sum-capacity of that 
MAC with a $(\mathbf{C}_\mathrm{in},\mathbf{C}_\mathrm{out})$-CF 
has an infinite derivative in the direction of every
$\mathbf{v}\in\mathbb{R}^k_{>0}$ at $\mathbf{C}_\mathrm{out}=\mathbf{0}$.
In other words, as a function of $\mathbf{C}_\mathrm{out}$, the 
sum-capacity grows faster than any function with bounded derivative at 
$\mathbf{C}_\mathrm{out}=\mathbf{0}$. This means that for any MAC in 
$\mathcal{C}^*$, sharing a small number of bits with each encoder leads to
a large gain in sum-capacity.  

An important implication of this result is the existence of a memoryless
network that does not satisfy the ``edge removal property'' 
\cite{HoEtAl, JalaliEtAl}. A network satisfies the edge removal property
if removing an edge of capacity $\delta>0$ changes the capacity region 
by at most $\delta$ in each dimension. Thus removing an edge of 
capacity $\delta$ from a network which has $k$ sources and
satisfies the edge removal property, decreases sum-capacity 
by at most $k\delta$, a linear function of $\delta$. Now consider a network
consisting of a MAC in $\mathcal{C}^*$ and a 
$(\mathbf{C}_\mathrm{in},\mathbf{C}_\mathrm{out})$-CF, where 
$\mathbf{C}_\mathrm{in}\in\mathbb{R}^k_{>0}$. Our main result 
(Theorem \ref{thm:sumCapacity}) implies that for small 
$\mathbf{C}_\mathrm{out}$, removing 
all the output edges reduces sum-capacity by an amount much larger than
$k\sum_{j\in [k]}C_\mathrm{out}^j$. Thus there exist memoryless networks
that do not satisfy the edge removal property. The first example of such a 
network appeared in \cite{NoorzadEtAl}. 

We introduce the coding scheme that leads to Theorem \ref{thm:sumCapacity}
in Section~\ref{sec:theCode}. This scheme combines 
forwarding, coordination, and classical MAC coding.  
In forwarding, each encoder sends part of its message to all other encoders 
by passing that information through the CF.\footnote{While it is 
possible to consider encoders that send \emph{different} parts of their messages
to different encoders using Han's result for the MAC with correlated sources~\cite{Han}, 
we avoid these cases for simplicity.} 
When $k=2$, forwarding is equivalent to a single round of conferencing as described
in \cite{WillemsMAC}. The coordination strategy
is a modified version of Marton's coding scheme for the broadcast channel \cite{Marton,ElGamalMeulen}.
To implement this strategy, the CF shares information with the encoders that enables them to 
transmit codewords that are jointly typical with respect to a {\em dependent} distribution; 
this is proven using a multivariate version of the covering lemma~\cite[p. 218]{ElGamalKim}.
The multivariate covering lemma is stated for strongly typical sets in \cite{ElGamalKim}. 
In Appendix \ref{app:MCL}, using the proof of the 2-user case from \cite{ElGamalKim}
and techniques from \cite{KoetterEtAl2}, we prove this lemma for weakly typical sets
\cite[p. 251]{CoverThomas}. Using weakly typical sets in our achievability proof 
allows our results to extend to continuous (e.g., Gaussian) channels without the 
need for quantization. Finally, the classical MAC strategy is Ulrey's \cite{Ulrey} 
extension of Ahlswede's \cite{Ahlswede1,Ahlswede2} and Liao's \cite{Liao} 
coding strategy to the $k$-user MAC. 

Using techniques from Willems \cite{WillemsMAC}, we derive an outer bound 
(Proposition \ref{prop:outerBound}) for the capacity region of the MAC 
with a $(\mathbf{C}_\mathrm{in},\mathbf{C}_\mathrm{out})$-CF. This outer bound 
does not capture the dependence of the capacity region on $\mathbf{C}_\mathrm{out}$
and is thus loose for some values of $\mathbf{C}_\mathrm{out}$. However, if the 
entries of $\mathbf{C}_\mathrm{out}$ are sufficiently larger than the entries of 
$\mathbf{C}_\mathrm{in}$, then our inner and outer bounds agree and we obtain 
the capacity region (Corollary \ref{cor:capacityRegion}).

In Section~\ref{sec:twoUser}, we apply our results to the 2-user Gaussian MAC 
with a CF that has access to the messages of both encoders and has links 
of output capacity $C_\mathrm{out}$. We show that for small 
$C_\mathrm{out}$, the achievable sum-rate approximately equals a constant times
$\sqrt{C_\mathrm{out}}$. A similar approximation holds for a weighted version of the 
sum-rate as well, as we see in Proposition \ref{prop:gaussianSlope}. This result implies
that at least for the 2-user Gaussian MAC, the benefit of cooperation is not limited to
sum-capacity and applies to other capacity region metrics as well. 

In Section~\ref{sec:conf}, we consider the extension of 
Willems' conferencing model~\cite{WillemsMAC} 
from 2 to $k$ users. A special case of this model with $k=3$
is studied in~\cite{SimeoneEtAl} for the Gaussian MAC. 
While the authors of \cite{SimeoneEtAl} use two conferencing rounds
in their achievability result, it is not clear from \cite{SimeoneEtAl}
if there is a benefit in using two rounds instead of one, and if so, how
large that benefit is. Here we explicitly show that a single conferencing round
is not optimal for $k\geq 3$, even though it is known to be optimal when $k=2$ 
\cite{WillemsMAC}. Finally, we apply our outer bound for the $k$-user MAC with a CF 
to obtain an outer bound for the $k$-user MAC with conferencing. 
The resulting outer bound is tight when $k=2$.

In the next section, we formally define the capacity region of the network
consisting of a $k$-user MAC and a CF. 

\section{Model} \label{sec:model}

Consider a network with $k$ encoders, a CF, a $k$-user MAC,
and a decoder (Figure \ref{fig:network}).  
For each $j\in[k]$, encoder $j$ communicates with the CF 
using noiseless links of capacities $C_\mathrm{in}^j\geq 0$ and $C_\mathrm{out}^j\geq 0$ 
going to and from the CF, respectively.
The $k$ encoders communicate with the decoder through 
a MAC $(\mathcal{X}_{[k]},p(y|x_{[k]}),\mathcal{Y})$, where 
\begin{equation*}
  \mathcal{X}_{[k]}=\prod_{j=1}^k \mathcal{X}_j,
\end{equation*}
and an element of $\mathcal{X}_{[k]}$ is denoted by $x_{[k]}$. 
We say a MAC is discrete if  
$\mathcal{X}_{[k]}$ and $\mathcal{Y}$ are either finite 
or countably infinite, and 
$p(y|x_{[k]})$ is a probability mass function on $\mathcal{Y}$ for
every $x_{[k]}\in\mathcal{X}_{[k]}$. We say a MAC is continuous if
$\mathcal{X}_{[k]}=\mathbb{R}^k$, $\mathcal{Y}=\mathbb{R}$, and 
$p(y|x_{[k]})$ is a probability density function on $\mathcal{Y}$ for 
all $x_{[k]}$.  
In addition, we assume that our channel is memoryless and 
without feedback \cite[p. 193]{CoverThomas}, so that for every positive integer 
$n$, the $n$th extension channel of our MAC is given by $p(y^n|x_{[k]}^n)$,
where
\begin{equation*}
  \forall
	(x_{[k]}^n,y^n)\in\mathcal{X}^n_{[k]}\times\mathcal{Y}^n:
  p(y^n|x_{[k]}^n)=\prod_{t=1}^n
	p(y_t|x_{[k]t}). 
\end{equation*}
An example of a continuous MAC is the $k$-user Gaussian MAC 
with noise variance $N>0$, where 
\begin{equation} \label{eq:mUserGaussian}
  p(y|x_{[k]})=\frac{1}{\sqrt{2\pi N}}
	\exp\Big[-\frac{1}{2N}\Big(y-\sum_{j\in [k]}x_j\Big)^2\Big]
\end{equation}
Henceforth, all MACs are memoryless and without feedback, and
either discrete or continuous. 

We next describe a
\begin{equation*}
  \big((2^{nR_1},\dots,2^{nR_k}),n,L\big)
	\text{-code}
\end{equation*}
for the MAC $(\mathcal{X}_{[k]},p(y|x_{[k]}),\mathcal{Y})$
with a $(\mathbf{C}_\mathrm{in},\mathbf{C}_\mathrm{out})$-CF
with cost functions $(b_j)_{j\in [k]}$ and cost constraint vector
$\mathbf{B}=(B_j)_{j\in [k]}\in \mathbb{R}^k_{\geq 0}$. For each
$j\in [k]$, cost function $b_j$ is a fixed mapping from $\mathcal{X}_j$
to $\mathbb{R}_{\geq 0}$. 
Each encoder $j\in[k]$ wishes to transmit a message 
$w_j\in [2^{nR_j}]$ to the decoder. 
This is accomplished by first exchanging information with the CF 
and then transmitting across the MAC.  
Communication with the CF occurs in $L$ rounds. 
For each $j\in [k]$ and $\ell\in[L]$, 
sets $\mathcal{U}_{j\ell}$ and $\mathcal{V}_{j\ell}$, 
respectively, describe the alphabets of symbols that 
encoder $j$ can send to and receive from the CF in round $\ell$. 
These alphabets satisfy the link capacity constraints
\begin{align}
\sum_{\ell=1}^L \log|\mathcal{U}_{j\ell}| &\leq nC_\mathrm{in}^j\notag\\
\sum_{\ell=1}^L \log|\mathcal{V}_{j\ell}| &\leq nC_\mathrm{out}^j.\label{eq:VjlCout}
\end{align}
The operation of encoder $j$ and the CF, respectively, in round $\ell$ are given by 
\begin{align*}
  \varphi_{j\ell} &: [2^{nR_j}]\times \mathcal{V}_j^{\ell-1}
	\rightarrow \mathcal{U}_{j\ell} \\
  \psi_{j\ell} &:\prod_{i=1}^k \mathcal{U}_i^\ell
  \rightarrow \mathcal{V}_{j\ell}.
  \end{align*}
where $\mathcal{U}_j^\ell=\prod_{\ell'=1}^\ell\mathcal{U}_{j\ell'}$
and $\mathcal{V}_j^\ell=\prod_{\ell'=1}^\ell\mathcal{V}_{j\ell'}$.
After its exchange with the CF, encoder $j$ applies a function 
\begin{equation*}
  f_j :[2^{nR_j}] \times \mathcal{V}_j^L
  \rightarrow \mathcal{X}_j^n,
\end{equation*}
to choose a codeword, which it transmits across the channel. 
In addition, every $x_j^n$ in the range of $f_j$ satisfies
\begin{equation*}
  \sum_{t=1}^n b_j(x_{jt})\leq nB_j. 
\end{equation*}
The decoder receives channel output $Y^n$ and applies 
\begin{equation*}
  g:\mathcal{Y}^n\rightarrow \prod_{j=1}^k[2^{nR_j}]
\end{equation*}
to obtain estimate $\hat{W}_{[k]}$ of the message vector $w_{[k]}$.

The encoders, CF, and decoder together define a
\begin{equation*}
  \big((2^{nR_1},\dots,2^{nR_k}),n,L\big)\text{-code}.
\end{equation*}
The average error probability of the code is 
$P_e^{(n)}=\pr\big\{g(Y^n)\neq W_{[k]}\big\}$,
where $W_{[k]}$ is the transmitted message vector and is 
uniformly distributed on $\prod_{j=1}^k[2^{nR_j}]$.
A rate vector $R_{[k]}=(R_1,\dots, R_k)$ is {\em achievable} if 
there exists a sequence of $\big((2^{nR_1},\dots,2^{nR_k}),n,L\big)$ codes with
$P_e^{(n)}\rightarrow 0$ as $n\rightarrow\infty$. The capacity region,
$\mathscr{C}(\mathbf{C}_\mathrm{in},\mathbf{C}_\mathrm{out})$, is defined
as the closure of the set of all achievable rate vectors.

\section{Results} \label{sec:results}

In this section, we describe the key results. 
In Subsection \ref{subsec:innerBound}, we present our inner 
bound. In Subsection \ref{subsec:coopBenefit}, we state our main
result, which proves the existence of a class of MACs with 
large cooperation gain. Finally, in Subsection \ref{subsec:outerBound},
we discuss our outer bound. 

\subsection{Inner Bound} \label{subsec:innerBound}

Using the coding scheme we introduce in Section \ref{sec:theCode}, 
we obtain an inner bound for the capacity region of the $k$-user MAC with a 
$(\mathbf{C}_\mathrm{in},\mathbf{C}_\mathrm{out})$-CF.
The following definitions are useful for describing that bound.
Choose vectors $\mathbf{C}_0=(C_{j0})_{j=1}^k$ and 
$\mathbf{C}_d=(C_{jd})_{j=1}^k$ in $\mathbb{R}^k_{\geq 0}$
such that for all $j\in [k]$,
\begin{align} 
  C_{j0} &\leq C_\mathrm{in}^j \label{eq:C0Cd1}\\
  C_{jd}+\sum_{i\neq j}C_{i0} &\leq C_\mathrm{out}^j.\label{eq:C0Cd2}
\end{align}
Here $C_{j0}$ is the number of bits per channel use 
encoder $j$ sends directly to the other encoders via
the CF and $C_{jd}$ is the number of bits per channel use
the CF transmits to encoder $j$ to implement the coordination strategy. 
Subscript ``$d$'' in $C_{jd}$ alludes to the dependence created through coordination.
Let $S_d=\big\{j\in [k]:C_{jd}\neq 0\big\}$ be the set of encoders 
that participate in this dependence.

Fix alphabets $\mathcal{U}_0,\mathcal{U}_1,\dots,\mathcal{U}_k$. For every 
nonempty $S\subseteq [k]$, let $\mathcal{U}_S$ be
the set of all $u_S=(u_j)_{j\in S}$ where $u_j\in\mathcal{U}_j$
for all $j\in S$. Define the set $\mathcal{X}_S$
similarly.
Let $\mathcal{P}(\mathcal{U}_0,\mathcal{U}_{[k]},\mathcal{X}_{[k]},S_d)$ 
be the set of all distributions
on $\mathcal{U}_0\times\mathcal{U}_{[k]}\times\mathcal{X}_{[k]}$ that
are of the form
\begin{equation} \label{eq:dist}
  p(u_0)\cdot\prod_{i\in S_d^c}p(u_i|u_0)
  \cdot p(u_{S_d}|u_0,u_{S_d^c})\cdot\prod_{j\in [k]} p(x_j|u_0,u_j),
\end{equation}
satisfy the dependence constraints\footnote{The constraint on $\zeta_S$ is imposed by the multivariate 
covering lemma (Appendix \ref{app:MCL}), which we use in the 
proof of our inner bound.}
\begin{equation*}
  \zeta_S:=\sum_{j\in S}C_{jd}-\sum_{j\in S}H(U_j|U_0)+H(U_S|U_0,U_{S_d^c})>0
	\qquad\forall\:\emptyset\subsetneq S\subseteq S_d,
\end{equation*} 
and cost constraints 
\begin{equation} \label{eq:cost}
  \mathbb{E}\big[b_j(X_j)\big]\leq B_j
	\qquad\forall j\in [k]. 
\end{equation}

Here $U_0$ encodes the ``common message,'' 
which, for every $j\in [k]$, contains $nC_{j0}$ bits from the message of encoder $j$ 
and is shared with all other encoders through the CF; each random variable 
$U_{j}$ captures the information encoder $j$ receives from the CF to
create dependence with the codewords of other encoders. The random variable
$X_j$ represents the symbol encoder $j$ transmits over the channel. 

For any $\mathbf{C}_0,\mathbf{C}_d\in\mathbb{R}^k_{\geq 0}$ 
satisfying (\ref{eq:C0Cd1}) and (\ref{eq:C0Cd2}) 
and any $p\in\mathcal{P}(\mathcal{U}_0,\mathcal{U}_{[k]},\mathcal{X}_{[k]},S_d)$, let 
$\mathscr{R}(\mathbf{C}_0,\mathbf{C}_d,p)$ be the set of all $(R_1,\dots,R_k)$ 
for which 
\begin{equation} \label{eq:thmIBsum}
  \sum_{j\in [k]} R_j <I(X_{[k]};Y)-\zeta_{S_d},
\end{equation} 
and for every $S,T\subseteq [k]$,  
\begin{align} 
  \MoveEqLeft
  \sum_{j\in A}(R_j-C_{j0})^++\sum_{j\in B\cap T}(R_j-C_\mathrm{in}^j)^+\notag \\
   &<I\big(U_A,X_{A\cup (B\cap T)};Y\big|U_0,U_B,X_{B\setminus T}
	\big)-\zeta_{(A\cup B)\cap S_d}\label{eq:thmIB}
\end{align}
holds for some sets $A$ and $B$ for which $S\cap S_d^c \subseteq A  \subseteq S$ 
and $S^c\cap S_d^c \subseteq B \subseteq S^c$.

We next state our inner bound for the $k$-user MAC with encoder cooperation
via a CF. The coding strategy that achieves this inner bound uses only
a single round of cooperation ($L=1$). 
The proof is given in Subsection \ref{subsec:IBproof}.

\begin{thm}[Inner Bound] \label{thm:innerBound}
For any MAC $(\mathcal{X}_{[k]},p(y|x_{[k]}),\mathcal{Y})$
with a $(\mathbf{C}_\mathrm{in},\mathbf{C}_\mathrm{out})$-CF,
\begin{equation*}
  \mathscr{C}(\mathbf{C}_\mathrm{in},\mathbf{C}_\mathrm{out})\supseteq
  \overline{\bigcup \mathscr{R}(\mathbf{C}_0,\mathbf{C}_d,p)}
\end{equation*}
where $\bar{A}$ denotes the closure of set $A$ 
and the union is over all $\mathbf{C}_0$ and $\mathbf{C}_d$ 
satisfying (\ref{eq:C0Cd1}) and (\ref{eq:C0Cd2}), and 
$p\in\mathcal{P}(\mathcal{U}_0,\mathcal{U}_{[k]},\mathcal{X}_{[k]},S_d)$.
\end{thm}  

The achievable region given in Theorem \ref{thm:innerBound} is convex
and thus we do not require the convex hull operation. The proof is
similar to \cite{CoverElGamal, Noorzad2} and is omitted. 

The next corollary 
treats the case where the CF transmits the bits it receives 
from each encoder to all other encoders without change.
In this case, our coding strategy simply combines forwarding with
classical MAC encoding. 
We obtain this result from Theorem \ref{thm:innerBound} 
by setting $C_{jd}=0$ and $|\mathcal{U}_j|=1$ for all $j\in [k]$ 
and choosing $A=S$ and $B=S^c$ for every $S,T\subseteq [k]$.
In Corollary~\ref{cor:forwardInnerBound}, 
$\mathcal{P}_\mathrm{ind}(\mathcal{U}_0,\mathcal{X}_{[k]})$
is the set of all distributions $p(u_0)\prod_{j\in [k]}p(x_j|u_0)$
that satisfy the cost constraints (\ref{eq:cost}). 

\begin{cor}[Forwarding Inner Bound] \label{cor:forwardInnerBound}
The capacity region of any MAC
with a $(\mathbf{C}_\mathrm{in},\mathbf{C}_\mathrm{out})$-CF
contains the set of all rate vectors that for some constants 
$(C_{j0})_{j\in [k]}$
(satisfying (\ref{eq:C0Cd1}) and (\ref{eq:C0Cd2})
with $C_{jd}=0$ for all $j$) and 
some distribution $p\in \mathcal{P}_\mathrm{ind}(\mathcal{U}_0,\mathcal{X}_{[k]})$,
satisfy
\begin{align*}
  \sum_{j\in S}R_j
  &<I\big(X_S;Y|U_0,X_{S^c})+\sum_{j\in S}C_{j0}
	\qquad\forall\:\emptyset\neq S\subseteq [k]\\
  \sum_{j\in [k]} R_j &<I(X_{[k]};Y).
\end{align*} 
\end{cor}  

\subsection{Sum-Capacity Gain} \label{subsec:coopBenefit}

We wish to understand when cooperation leads to a benefit that 
exceeds the resources employed to enable it. 
Therefore, we compare the gain in sum-capacity obtained through 
cooperation to the number of bits shared with the encoders to 
enable that gain. 

For any $k$-user MAC with 
a $(\mathbf{C}_\mathrm{in},\mathbf{C}_\mathrm{out})$-CF, define
the sum-capacity as
\begin{equation*}
  C_\mathrm{sum}(\mathbf{C}_\mathrm{in},\mathbf{C}_\mathrm{out})
  =\max_{\mathscr{C}(\mathbf{C}_\mathrm{in},\mathbf{C}_\mathrm{out})}
	\sum_{j=1}^k R_j.
\end{equation*}
For a fixed $\mathbf{C}_\mathrm{in}\in\mathbb{R}^k_{\geq 0}$, define the
``sum-capacity gain''  $G:\mathbb{R}^k_{\geq 0}\rightarrow \mathbb{R}_{\geq 0}$
as 
\begin{equation*}
  G(\mathbf{C}_\mathrm{out})=C_\mathrm{sum}
	(\mathbf{C}_\mathrm{in},\mathbf{C}_\mathrm{out})
	-C_\mathrm{sum}(\mathbf{C}_\mathrm{in},\mathbf{0}),
\end{equation*}
where $\mathbf{C}_\mathrm{out}=(C^j_\mathrm{out})_{j=1}^k$
and $\mathbf{0}=(0,\dots,0)$.
Note that regardless of $\mathbf{C}_\mathrm{in}$, 
it follows from (\ref{eq:VjlCout}) that 
no cooperation is possible
when $\mathbf{C}_\mathrm{out}=\mathbf{0}$. Thus 
\begin{equation*}
  C_\mathrm{sum}(\mathbf{C}_\mathrm{in},\mathbf{0})=
  C_\mathrm{sum}(\mathbf{0},\mathbf{0})=
	\max_{p\in\mathcal{P}_\mathrm{ind}(X_{[k]})}I(X_{[k]};Y),
\end{equation*}
where $\mathcal{P}_\mathrm{ind}(\mathcal{X}_{[k]})$ is the set of all independent 
distributions
\begin{equation*}
  p(x_{[k]})=\prod_{j\in [k]}p(x_j)
\end{equation*}
on $\mathcal{X}_{[k]}$ that satisfy the cost constraints $(\ref{eq:cost})$. 
Similarly, $\mathcal{P}(\mathcal{X}_{[k]})$ is the set of 
\emph{all} distributions on 
$\mathcal{X}_{[k]}$ that satisfy  $(\ref{eq:cost})$.

For sets $\mathcal{X}_1,\dots,\mathcal{X}_k,\mathcal{Y}$, cost functions $(b_j)_{j\in [k]}$,
and cost constraints $(B_j)_{j\in [k]}$, we next define a special class of 
MACs $\mathcal{C}^*(\mathcal{X}_{[k]},\mathcal{Y})$. 
We say a MAC $(\mathcal{X}_{[k]},p(y|x_{[k]}),\mathcal{Y})$
is in $\mathcal{C}^*(\mathcal{X}_{[k]},\mathcal{Y})$, if there exists
$p_\mathrm{ind}\in \mathcal{P}_\mathrm{ind}(X_{[k]})$ that satisfies
\begin{equation*}
	I_\mathrm{ind}(X_{[k]};Y)=
	\max_{p\in \mathcal{P}_\mathrm{ind}(X_{[k]})}I(X_{[k]};Y),
\end{equation*}
and $p_\mathrm{dep}\in \mathcal{P}(X_{[k]})$ whose support is contained
in the support of $p_\mathrm{ind}$ and satisfies
\begin{equation} \label{eq:depBiggerInd}
  I_\mathrm{dep}(X_{[k]};Y)+D\big(p_\mathrm{dep}(y)\|p_\mathrm{ind}(y)\big)
	> I_\mathrm{ind}(X_{[k]};Y).
\end{equation}
In the above equation, $p_\mathrm{dep}(y)$ and $p_\mathrm{ind}(y)$ are 
the output distributions corresponding to the input distributions
$p_\mathrm{dep}(x_{[k]})$ and $p_\mathrm{ind}(x_{[k]})$, respectively. 
We remark that (\ref{eq:depBiggerInd}) is equivalent to 
\begin{equation*}
  \mathbb{E}_\mathrm{dep}\Big[D\big(p(y|X_{[k]})\|p_\mathrm{ind}(y)\big)\Big]
	>\mathbb{E}_\mathrm{ind}\Big[D\big(p(y|X_{[k]})\|p_\mathrm{ind}(y)\big)\Big],
\end{equation*}
where the expectations are with respect to $p_\mathrm{dep}(x_{[k]})$
and $p_\mathrm{ind}(x_{[k]})$, respectively. 

Using these definitions, we state our main result which captures a family
of MACs for which the slope of the gain function is infinite in every direction
at $\mathbf{C}_\mathrm{out}=\mathbf{0}$. In this statement, 
for any unit vector $\mathbf{v}\in\mathbb{R}^k_{\geq 0}$,
$D_\mathbf{v}G$ is the directional derivative of $G$ in the 
direction of $\mathbf{v}$. The proof appears in 
Subsection \ref{subsec:sumCapacity}. 

\begin{thm}[Sum-capacity] \label{thm:sumCapacity}
Let  $(\mathcal{X}_{[k]},p(y|x_{[k]}),\mathcal{Y})$ be a MAC in 
$\mathcal{C}^*(\mathcal{X}_{[k]},\mathcal{Y})$ and
$\mathbf{C}_\mathrm{in}\in \mathbb{R}_{>0}^k$.
Then for any unit vector $\mathbf{v}\in\mathbb{R}^k_{>0}$,
\begin{equation*}
 (D_{\mathbf{v}}G)(\mathbf{0})=\infty.
\end{equation*}
\end{thm} 

Note that for continuous MACs, when for $j\in [k]$ and 
$x\in\mathbb{R}$, $b_j(x)=x^2$, cost constraints are referred
to as power constraints. In addition, for every $j\in [k]$, 
the variable $P_j$ is commonly used instead of $B_j$. Our next 
proposition provides necessary and sufficient conditions
under which the $k$-user Gaussian MAC with power constraints
is in $\mathcal{C}^*(\mathbb{R}^k,\mathbb{R})$. The proof
is provided in Subsection \ref{subsec:kUserGaussian}. 
\begin{prop} \label{prop:kUserGaussian}
The $k$-user Gaussian MAC with power constraint vector 
$\mathbf{P}=(P_j)_{j\in [k]}\in\mathbb{R}^k_{\geq 0}$ is 
in $\mathcal{C}^*(\mathbb{R}^k,\mathbb{R})$ if and only 
if at least two entries of $\mathbf{P}$ are positive.
\end{prop}

\subsection{Outer Bound} \label{subsec:outerBound}

We next describe our outer bound. While we only make use of a single 
round of cooperation in our inner bound (Theorem \ref{thm:innerBound}), 
the outer bound applies to all coding schemes regardless of the
number of rounds.

\begin{prop}[Outer Bound] \label{prop:outerBound}
For the MAC 
$(\mathcal{X}_{[k]},p(y|x_{[k]}),\mathcal{Y})$,
$\mathscr{C}(\mathbf{C}_\mathrm{in},\mathbf{C}_\mathrm{out})$
is a subset of the set of all rate vectors that 
for some distribution 
$p\in\mathcal{P}_\mathrm{ind}(\mathcal{U}_0,\mathcal{X}_{[k]})$
satisfy
\begin{align} 
  \sum_{j\in S}R_j &\leq I\big(X_S;Y|U_0,X_{S^c}\big)+\sum_{j\in S}C^j_\mathrm{in}
	\qquad\forall\:\emptyset\neq S\subseteq [k]\label{eq:thmOB}\\
  \sum_{j\in [k]} R_j &\leq I(X_{[k]};Y).\label{eq:thmOBsum}
\end{align} 
\end{prop}  

The proof of this proposition is given in Subsection \ref{subsec:outerBoundProof}. Our
proof uses ideas similar to the proof of the converse for the 2-user MAC with 
conferencing \cite{WillemsMAC}. 

If the capacities of the CF output links are sufficiently large,
our inner and outer bounds coincide and we obtain the capacity region.
This follows by setting $C_{j0}=C_\mathrm{in}^j$ for all $j\in [k]$
in our forwarding inner bound (Corollary \ref{cor:forwardInnerBound}) and comparing it 
with the outer bound given in Proposition \ref{prop:outerBound}.  
\begin{cor} \label{cor:capacityRegion}
For the MAC $(\mathcal{X}_{[k]},p(y|x_{[k]}),\mathcal{Y})$
with a $(\mathbf{C}_\mathrm{in},\mathbf{C}_\mathrm{out})$-CF, if
\begin{equation*}
  \forall j\in [k]:
  C^j_\mathrm{out}\geq \sum_{i:i\neq j}C^i_\mathrm{in},
\end{equation*}
then our inner and outer bounds agree. 
\end{cor}

\section{The Coding Scheme} \label{sec:theCode}

Choose nonnegative constants $(C_{j0})_{j=1}^k$ and $(C_{jd})_{j=1}^k$
such that (\ref{eq:C0Cd1})
and (\ref{eq:C0Cd2}) hold for all $j\in [k]$. 
Fix a distribution $p\in\mathcal{P}(\mathcal{U}_0,\mathcal{U}_{[k]},\mathcal{X}_{[k]},S_d)$
and constants $\epsilon,\delta>0$. Let 
\begin{align*}
  R_{j0} &= \min\{R_j,C_{j0}\}\\
  R_{jd} &= \min\{R_j,C_\mathrm{in}^j\}-R_{j0}\\
  R_{jj} &= R_j-R_{j0}-R_{jd}=(R_j-C_\mathrm{in}^j)^+,
\end{align*}
where $x^+=\max\{x,0\}$ for any real number $x$. For every $j\in [k]$,
split the message of encoder $j$ as $w_j=(w_{j0},w_{jd},w_{jj})$, where 
$w_{j0}\in [2^{nR_{j0}}]$, $w_{jd}\in [2^{nR_{jd}}]$, $w_{jj}\in [2^{nR_{jj}}]$.
For all $j\in [k]$, encoder $j$ sends $(w_{j0},w_{jd})$ noiselessly to the CF. 
This is possible, since $R_{j0}+R_{jd}$ is less than or equal to 
$C_\mathrm{in}^j$. The CF sends $w_{j0}$ to all other encoders via its output links
and uses $w_{jd}$ to implement the coordination strategy to be descibed
below. Due to the CF rate constraints, encoder $j$ cannot share the remaining 
part of its message, $w_{jj}$, with the CF. Instead, 
it transmits $w_{jj}$ over the channel
using the classical MAC strategy. 

Let $\mathcal{W}_0 = \prod_{j=1}^k [2^{nR_{j0}}]$.
For every $w_0\in \mathcal{W}_0$, let $U_0^n(w_0)$ be 
drawn independently according to 
\begin{equation*}
  \pr\big\{U_0^n(w_0)=u_0^n\big\}=
  \prod_{t=1}^n p(u_{0t}).
\end{equation*}
Given $U_0^n(w_0)=u_0^n$, for every $j\in [k]$, $w_{jd}\in [2^{nR_{jd}}]$, and
$z_j\in [2^{nC_{jd}}]$, let $U_j^n(w_{jd},z_j|u_0^n)$ be drawn independently 
according to 
\begin{equation} \label{eq:mu1}
  \pr\Big\{U_j^n(w_{jd},z_j|u_0^n)=u_j^n\Big|U_0^n(w_0)=u_0^n\Big\}
  =\prod_{t=1}^n p(u_{jt}|u_{0t}).
\end{equation}
For every $(w_1,\dots,w_k)$, define 
$E(u_0^n,\mu_1,\dots,\mu_k)$ as the event where
$U_0^n(w_0)=u_0^n$ and for every $j\in [k]$,
\begin{equation} \label{eq:mu2}
  U_j^n(w_{jd},\cdot|u_0^n)=\mu_j(\cdot),
\end{equation}
where $\mu_j$ is a mapping from $[2^{nC_{jd}}]$ to $\mathcal{U}_j^n$.
Let $\mathcal{A}(u_0^n,\mu_{[k]})$ be the set of
all $z_{[k]}=(z_1,\dots,z_k)$ such that 
\begin{equation} \label{eq:mu3}
  \big(u_0^n,\mu_{[k]}(z_{[k]})\big)
  \in A_\delta^{(n)}(U_0,U_{[k]}),
\end{equation}
where $\mu_{[k]}(z_{[k]})=(\mu_1(z_1),\ldots,\mu_k(z_k))$ 
and $A_\delta^{(n)}(U_0,U_{[k]})$ is the weakly typical 
set with respect to the distribution $p(u_0,u_{[k]})$.
If $\mathcal{A}(u_0^n,\mu_{[k]})$ is empty,
set $Z_j=1$ for all $j\in [k]$. Otherwise, let the
$k$-tuple $Z_{[k]}=(Z_1,\dots,Z_k)$ be the smallest 
element of $\mathcal{A}(u_0^n,\mu_{[k]})$ with respect to 
the lexicographical order. 
Finally, given $U_0^n(w_0)=u_0^n$ and $U_j^n(w_{jd},Z_j|u_0^n)=u_j^n$, 
for each $w_{jj}\in [2^{nR_{jj}}]$, let
$X_j^n(w_{jj}|u_0^n,u_j^n)$ be a random vector drawn independently 
according to 
\begin{align*}
  \MoveEqLeft
  \pr\Big\{X_j^n(w_{jj}|u_0^n,u_j^n)=x_j^n\Big|U_0^n(w_0)=u_0^n,
  U_j^n(w_{jd},Z_j)=u_j^n\Big\}\\
  &=\prod_{j=1}^n p(x_{jt}|u_{0t},u_{jt}).
\end{align*}
We next describe the encoding and decoding processes. 

\textbf{Encoding.} 
For every $j\in [k]$, encoder $j$ sends the pair $(w_{j0},w_{jd})$ to the
CF. The CF sends $((w_{i0})_{i\neq j},Z_j)$ back to encoder $j$. 
Encoder $j$, having access to $w_0=(w_{j0})_j$ and $Z_j$, 
transmits $X_j^n(w_{jj}|U_0^n(w_0),U_j^n(w_{jd},Z_j))$
over the channel. 

\textbf{Decoding.}
The decoder, upon receiving $Y^n$, 
maps $Y^n$ to the unique $k$-tuple $\hat{W}_{[k]}$ such that
\begin{align} \label{eq:dec}
  \MoveEqLeft
  \Big(U_0^n(\hat{W}_0),\big(U_j^n(\hat{W}_{jd},\hat{Z}_j|U_0^n)\big)_j,
  \big(X_j^n(\hat{W}_{jj}|U_0^n,U_j^n)\big)_j,Y^n\Big)\notag\\
  &\in A_\epsilon^{(n)}(U_0,U_{[k]},X_{[k]},Y).
\end{align}
If such a $k$-tuple does not exist, the decoder sets its 
output to the $k$-tuple $(1,1,\dots,1)$. 

The analysis of the expected error probability for the proposed random code
appears in Subsection \ref{subsec:IBproof}.

\section{Case Study: 2-User Gaussian MAC} \label{sec:twoUser}
In this section, we study the network consisting of the 2-user Gaussian 
MAC with power constraints and a CF whose input link capacities are sufficiently large so that the
CF has full access to the messages and output link capacities both equal
$C_\mathrm{out}$. We show that in this scenario, the benefit of cooperation 
extends beyond sum-capacity; that is, capacity metrics other than sum-capacity
also exhibit an infinite slope at $C_\mathrm{out}=0$. In addition, we 
show that the behavior of these metrics (including sum-capacity) is bounded from
below by a constant multiplied $\sqrt{C_\mathrm{out}}$. 

From Theorem \ref{thm:innerBound},
it follows that the capacity region of our network contains the set
of all rate pairs $(R_1,R_2)$ that satisfy
\begin{align*}
 R_1 &\leq \max\{I(X_1;Y|U_0)-C_{1d},I(X_1;Y|X_2,U_0)-\zeta\}+C_{10}\\
 R_2 &\leq \max\{I(X_2;Y|U_0)-C_{2d},I(X_2;Y|X_1,U_0)-\zeta\}+C_{20}\\
 R_1+R_2 &\leq I(X_1,X_2;Y|U_0)-\zeta+C_{10}+C_{20}\\
 R_1+R_2 &\leq I(X_1,X_2;Y)-\zeta
\end{align*}
for some nonnegative constants $C_{1d},C_{2d}\leq C_\mathrm{out}$,
\begin{align*}
 C_{10} &= C_\mathrm{out}-C_{2d}\\
 C_{20} &= C_\mathrm{out}-C_{1d},
\end{align*}
and some distribution $p(u_0)p(x_1,x_2|u_0)$ that satisfies 
$\mathbb{E}[X_i^2]\leq P_i$ for $i\in\{1,2\}$ and 
\begin{equation*}
  \zeta := C_{1d}+C_{2d}-I(X_1;X_2|U_0)\geq 0.
\end{equation*}

By (\ref{eq:mUserGaussian}), the 2-user Gaussian MAC can be
represented as
\begin{equation*}
  Y = X_1+X_2 + Z,
\end{equation*}
where $Z$ is independent of $(X_1,X_2)$, and is distributed as 
$Z\sim\mathcal{N}(0,N)$ for some noise variance $N>0$. 
Let $U_0\sim \mathcal{N}(0,1)$, and $(X'_1,X'_2)$ be a pair 
of random variables independent of $U_0$ and jointly distributed as 
$\mathcal{N}(\mu,\Sigma)$, where
\begin{equation*}
  \mu=\begin{pmatrix}
  0 \\0
  \end{pmatrix},
  \Sigma=
  \begin{pmatrix}
  1 & \rho_0 \\
  \rho_0 & 1
 \end{pmatrix}
\end{equation*}
for some $\rho_0\in [0,1]$. Finally, for $i\in\{1,2\}$, set 
\begin{equation*}
  \frac{1}{\sqrt{P_i}}X_i=\rho_i X'_i
	+\sqrt{1-\rho_i^2}U_0,
\end{equation*}
for some $\rho_i\in [0,1]$. Calculating the region described above for 
the Gaussian MAC using the joint
distribution of $(U_0,X_1,X_2)$ and setting $\gamma_i=P_i/N$ for $i\in\{1,2\}$ and 
$\bar{\gamma}=\sqrt{\gamma_1\gamma_2}$, gives the set of all rate pairs 
$(R_1,R_2)$ satisfying
\begin{align*} 
R_1 &\leq \max\bigg\{\frac{1}{2}\log
\frac{1+\rho_1^2\gamma_1+\rho_2^2\gamma_2+2\rho_0\rho_1\rho_2\bar{\gamma}}
{1+(1-\rho_0^2)\rho_2^2\gamma_2}-C_{1d},
\frac{1}{2}\log\big(1+(1-\rho_0^2)\rho_1^2\gamma_1\big)-\zeta\bigg\}+C_{10}\\
R_2 &\leq \max\bigg\{\frac{1}{2}\log
\frac{1+\rho_1^2\gamma_1+\rho_2^2\gamma_2+2\rho_0\rho_1\rho_2\bar{\gamma}}
{1+(1-\rho_0^2)\rho_1^2\gamma_1}-C_{2d},
\frac{1}{2}\log\big(1+(1-\rho_0^2)\rho_2^2\gamma_2\big)-\zeta\bigg\}+C_{20}
\end{align*}
and
\begin{align*}
R_1+R_2 &\leq \frac{1}{2}\log\big(1+\rho_1^2\gamma_1+\rho_2^2\gamma_2
+2\rho_0\rho_1\rho_2\bar{\gamma}\big)-\zeta+C_{10}+C_{20}\\
R_1+R_2 &\leq \frac{1}{2}\log\Big(1+\gamma_1+\gamma_2+2\big(\rho_0\rho_1\rho_2
+\sqrt{(1-\rho_1^2)(1-\rho_2^2)}\big)\bar{\gamma}\Big)-\zeta
\end{align*}
for some $\rho_1,\rho_2\in [0,1]$, and $0\leq \rho_0\leq \sqrt{1-2^{-2(C_{1d}+C_{2d})}}$.
Denote this region with $\mathscr{C}_\mathrm{ach}(C_\mathrm{out})$.

We next introduce a lower bound for the weighted version of the sum-capacity. 
Denote the capacity region of this network with $\mathscr{C}(C_\mathrm{out})$.
For every $\alpha\in [0,1]$, define 
\begin{equation*}
  C_\alpha(C_\mathrm{out})
	=\max_{(R_1,R_2)\in\mathscr{C}(C_\mathrm{out})}(\alpha R_1+(1-\alpha)R_2)
\end{equation*}
Note that $C_\alpha(C_\mathrm{out})$ is a generalization of the notion of
sum-capacity where the weighted sum of the encoders' rates is considered. 
The main result of this section demonstrates that for small $C_\mathrm{out}$,
$C_\alpha(C_\mathrm{out})$ is bounded from below by a constant times
$\sqrt{C_\mathrm{out}}$ when $C_\mathrm{out}$ is small. 
The proof is given in Subsection \ref{subsec:gaussianSlope}.
\begin{prop} \label{prop:gaussianSlope}
For the Gaussian MAC $Y=X_1+X_2+Z$ with $Z\sim\mathcal{N}(0,N)$ and input 
SNRs $(\gamma_1,\gamma_2)$, we have
\begin{equation*}
  C_\alpha(C_\mathrm{out})-C_\alpha(0)\geq 
	\frac{2\sqrt{\gamma_1\gamma_2\cdot\log e}}{1+\gamma_1+\gamma_2}
	\cdot\min\{\alpha,1-\alpha\}
	\cdot\sqrt{C_\mathrm{out}}+o(\sqrt{C_\mathrm{out}}).
\end{equation*}
In particular, for every $\alpha\in (0,1)$, 
\begin{equation*}
  \frac{dC_\alpha}{dC_\mathrm{out}}\Big|_{C_\mathrm{out}=0^+}=\infty.
\end{equation*}
\end{prop}

\begin{figure} 
\begin{center}
	\includegraphics[scale=0.35]{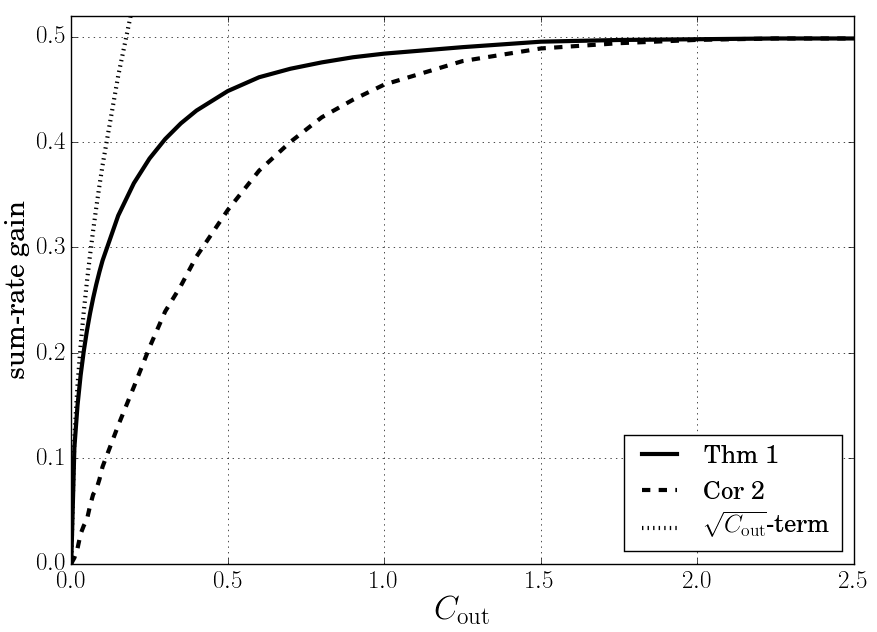}
	\caption{Plot of the achievable sum-rate gain given by 
	Theorem \ref{thm:innerBound} and Corollary \ref{cor:forwardInnerBound}
	for Gaussian input distributions, and
	the $\sqrt{C_\mathrm{out}}$-term given in Proposition \ref{prop:gaussianSlope}.
  Here $\gamma_1=\gamma_2=100$.} 
	\label{fig:plot}
\end{center}
\end{figure}

In Figure \ref{fig:plot}, using \cite{Hunter},
we plot the sum-rate of the region $\mathscr{C}_\mathrm{ach}(C_\mathrm{out})$
and the forwarding inner bound (Corollary \ref{cor:forwardInnerBound}) 
for $\gamma_1=\gamma_2=100$. We also plot the 
$\sqrt{C_\mathrm{out}}$-term in the lower bound given by Proposition
\ref{prop:gaussianSlope}. Notice that the
forwarding inner bound provides a cooperation gain that is at most linear
in $C_\mathrm{out}$.  

\section{The \texorpdfstring{$k$}{k}-User MAC with Conferencing Encoders} \label{sec:conf}

In this section, we extend Willems' conferencing encoders model 
\cite{WillemsMAC} from the 2-user MAC to the $k$-user MAC and 
provide an outer bound on the capacity region. 

\begin{figure} 
	\begin{center}
		\includegraphics[scale=0.35]{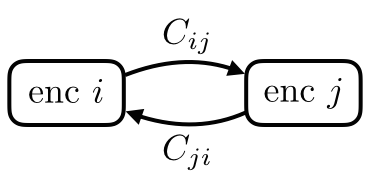}
		\caption{In $k$-user MAC with conferencing, for every
		$i,j\in [k]$, there are links of capacities $C_{ij}$ and 
		$C_{ji}$ connecting encoders $i$ and $j$.} \label{fig:conf}
	\end{center}
\end{figure}

Consider a $k$-user MAC where for every $i,j\in [k]$ 
(in this section, $i\neq j$ by assumption), 
there is a noiseless link of capacity $C_{ij}\geq 0$ going
from encoder $i$ to encoder $j$ and a noiseless link of 
capacity $C_{ji}\geq 0$ going back (Figure \ref{fig:conf}).
As in 2-user conferencing, the ``conference'' occurs over a finite number of rounds. 
In the first round, for every $i,j\in [k]$ with $C_{ij}>0$,
encoder $i$ transmits some information to encoder $j$ 
that is a function of its own message $w_i\in [2^{nR_i}]$. In each 
subsequent round, every encoder transmits information that is
a function of its message and information it receives before that round. 
Once the conference is over,
each encoder transmits its codeword over the $k$-user MAC. 

We next define a $\big((2^{nR_1},\dots,2^{nR_k}),n,L\big)$-code
for the $k$-user MAC with an $L$-round $(C_{ij})_{i,j=1}^k$-conference.
For every $i,j\in [k]$ and $\ell\in [L]$, fix a set $\mathcal{V}_{ij}^{(\ell)}$
so that for every $i,j\in [k]$, 
$\sum_{\ell=1}^L \log |\mathcal{V}_{ij}^{(\ell)}|\leq nC_{ij}$.
Here $\mathcal{V}_{ij}^{(\ell)}$ represents the alphabet of the symbol
encoder $i$ sends to encoder $j$ in round $\ell$ of the conference. 
For every $\ell\in [L]$, define 
$\mathcal{V}_{ij}^{\ell}=\prod_{\ell'=1}^\ell \mathcal{V}_{ij}^{(\ell')}$.
For $j\in [k]$, encoder $j$ is represented by the collection of functions
$\big(f_j,(h_{ji}^{(\ell)})_{i,\ell}\big)$
where 
\begin{align*}
  f_j&: [2^{nR_j}]\times \prod_{i:i\neq j}\mathcal{V}_{ij}^L
	\rightarrow \mathcal{X}_j^n\\
  h_{ji}^{(\ell)}&:[2^{nR_j}]\times\prod_{i':i'\neq j}\mathcal{V}_{i'j}^{\ell-1}
	\rightarrow \mathcal{V}_{ji}^{(\ell)}
\end{align*}
The decoder is a mapping
$g:\mathcal{Y}^n\rightarrow \prod_{j=1}^k [2^{nR_j}]$.
The definitions of cost constraints, achievable rate vectors, and the capacity region are
similar to those given in Section~\ref{sec:model}.

The next result compares the capacity region of a MAC with cooperation
under the conferencing and CF models. The proof is given in 
Subsection \ref{subsec:confCF}.

\begin{prop}\label{prop:confCF} 
The capacity region of a MAC with an $L$-round 
$(C_{ij})_{i,j=1}^k$-conference is a subset of the capacity region of the
same MAC with an $L$-round 
$(\mathbf{C}_\mathrm{in},\mathbf{C}_\mathrm{out})$-CF cooperation
if for all $j\in [k]$,
\begin{eqnarray*}
  C_\mathrm{in}^j\geq \sum_{i:i\neq j} C_{ji} & \mbox{ and }& 
	C_\mathrm{out}^j \geq \sum_{i:i\neq j} C_{ij}.
\end{eqnarray*}

Similarly, for every $L$, the capacity region of a MAC with $L$-round 
$(\mathbf{C}_\mathrm{in},\mathbf{C}_\mathrm{out})$-CF cooperation 
is a subset of the 
capacity region of the same MAC with a single-round $(C_{ij})_{i,j=1}^k$-conference
if for all $i,j\in [k]$, $C_{ij}\geq C_\mathrm{in}^i$.
\end{prop}

Combining the first part of Proposition~\ref{prop:confCF} 
with the outer bound from Proposition~\ref{prop:outerBound} 
results in the next corollary, which holds regardless of the 
number of conferencing rounds. 

\begin{cor}[Conferencing Outer Bound] 
The capacity region of a MAC with a $(C_{ij})_{i,j=1}^k$-conference 
is a subset of the set of all rate vectors $(R_1,\dots,R_k)$ that 
for some distribution 
$p\in\mathcal{P}_\mathrm{ind}(\mathcal{U}_0,\mathcal{X}_{[k]})$, 
satisfy
\begin{align*}
  \sum_{j\in S}R_j &\leq I\big(X_S;Y|U_0,X_{S^c}\big)
	+\sum_{j\in S}\sum_{i\neq j}C_{ji}\qquad
	\forall\:\emptyset\neq S\subseteq [k]\\
  \sum_{j\in [k]} R_j &\leq I(X_{[k]};Y).
\end{align*} 
\end{cor}  

While $k$-user conferencing is a direct extension of 2-user conferencing, there
is nonetheless an important difference when $k\geq 3$. While a single 
conferencing round suffices to achieve the capacity region in the 2-user case
\cite{WillemsMAC}, the same is not true when $k\geq 3$, as we next see. 

\begin{figure} 
	\begin{center}
		\includegraphics[scale=0.27]{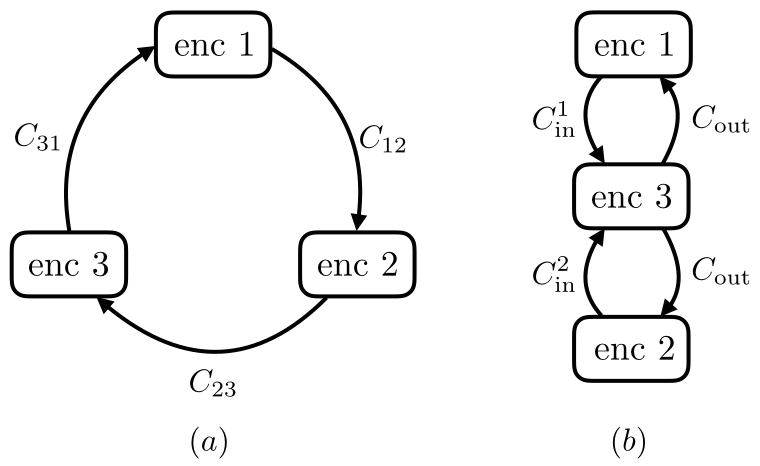}
		\caption{(a) The conferencing structure studied in \cite{SimeoneEtAl}. 
		(b) An example of a structure where allowing two conferencing rounds
		leads to a substantial gain over a single round.} \label{fig:twostruct}
	\end{center}
\end{figure}

A special case of this model for the 3-user Gaussian MAC, 
depicted in  Figure \ref{fig:twostruct}(a), is studied in \cite{SimeoneEtAl}. 
While the achievability scheme in \cite{SimeoneEtAl} uses two 
conferencing rounds, the magnitude of the gain resulting from using
an additional conferencing round is not clear. Here, using the idea 
of a cooperation facilitator, 
we consider an alternative shown in Figure \ref{fig:twostruct}(b), where
we show the possibility of a large cooperation gain when conferencing 
occurs in two rounds rather than one. 
Consider a 3-user MAC with conferencing.  
Fix positive constants $C_\mathrm{in}^1$ and $C_\mathrm{in}^2$. Let 
$C_{13}=C^1_\mathrm{in}$, $C_{23}=C^2_\mathrm{in}$,
$C_{31}=C_{32}=C_\mathrm{out}$ for 
$C_\mathrm{out}\in\mathbb{R}_{\geq 0}$, and $C_{12}=C_{21}=0$. 
Let $\mathscr{C}_1(C_\mathrm{out})$
and $\mathscr{C}_2(C_\mathrm{out})$ denote the capacity region of this network
with one and two rounds of conferencing, respectively. For each $L\in\{1,2\}$,
define the function $g_L(C_\mathrm{out})$ as
\begin{equation*}
  g_L(C_\mathrm{out})
  = \max_{(R_1,R_2,0)\in\mathscr{C}_L(C_\mathrm{out})}(R_1+R_2).
\end{equation*}
Note that when $L=1$, we have $g_1(C_\mathrm{out})=g_1(0)$ for all 
$C_\mathrm{out}\geq 0$,
since no cooperation is possible when encoder 3 is transmitting at rate zero. On the 
other hand, we next show that at least for some MACs, 
$g'_2(0)=\infty$; that is, $g_2$ has an infinite slope at $C_\mathrm{out}=0$.
Note that 
\begin{equation*}
  g_2(0)=g_1(0)=\max_{p(x_1)p(x_2),x_3}I(X_1,X_2;Y|X_3=x_3).
\end{equation*}
Suppose $x_3^*$ satisfies
\begin{equation*}
  \max_{p(x_1)p(x_2),x_3}I(X_1,X_2;Y|X_3=x_3)
	=\max_{p(x_1)p(x_2)}I(X_1,X_2;Y|X_3=x_3^*).
\end{equation*}
If the MAC $(\mathcal{X}_1\times\mathcal{X}_2,p(y|x_1,x_2,x_3^*),\mathcal{Y})$
is in $\mathcal{C}^*(\mathcal{X}_1\times\mathcal{X}_2,\mathcal{Y})$,
then by Theorem \ref{thm:sumCapacity}, we have $g'_2(0)=\infty$. 
Since $g_1$ is constant for all $C_\mathrm{out}$, 
while $g_2$ has an infinite slope at $C_\mathrm{out}=0$, and $g_1(0)=g_2(0)$, 
the two-round conferencing region is strictly larger than the single-round conferencing
region. Using the same technique, we can show a similar result for any $k\geq 3$;
that is, there exist $k$-user MACs where the two-round conferencing region
strictly contains the single-round region. 

\section{Proofs}

\subsection{Theorem \ref{thm:innerBound} (Inner bound)} \label{subsec:IBproof}

Fix $\eta>0$, and choose a distribution 
$p(u_0,u_{[k]},x_{[k]})$
on $\mathcal{U}_0\times\mathcal{U}_{[k]}\times\mathcal{X}_{[k]}$ of the form
\begin{equation*} 
  p(u_0)\cdot\prod_{i\in S_d^c}p(u_i|u_0)
  \cdot p(u_{S_d}|u_0,u_{S_d^c})\cdot\prod_{j\in [k]} p(x_j|u_0,u_j),
\end{equation*}
that satisfies the dependence constraints
\begin{equation*}
  \zeta_S:=\sum_{j\in S}C_{jd}-\sum_{j\in S}H(U_j|U_0)+H(U_S|U_0,U_{S_d^c})>0
	\qquad\forall\:\emptyset\subsetneq S\subseteq S_d,
\end{equation*} 
and cost constraints 
\begin{equation} \label{eq:costEta}
  \mathbb{E}\big[b_j(X_j)\big]\leq B_j-\eta
	\qquad\forall j\in [k]. 
\end{equation}

Let $(w_1,\dots,w_k)$ denote the transmitted $k$-tuple of messages
and $(\hat{W}_1,\dots,\hat{W}_k)$ denote the output of the decoder. 
To simplify notation, denote
\begin{equation*}
  U_0^n(w_0),U_j^n(w_{jd},Z_j|U_0^n), X_j^n(w_{jj}|U_0^n,U_j^n) 
\end{equation*}
with $U_0^n$, $U_j^n$, and $X_j^n$, respectively. Similarly, define
$\hat{U}_0^n$, $\hat{U}_j^n$, and $\hat{X}_j^n$ as
\begin{equation*}
  U_0^n(\hat{W}_0),U_j^n(\hat{W}_{jd},Z_j|U_0^n), X_j^n(\hat{W}_{jj}|U_0^n,U_j^n).
\end{equation*}
Here $\hat{W}_0$, $\hat{W}_{jd}$, and $\hat{W}_{jj}$ are defined in terms of 
$(\hat{W}_j)_j$ similar to the definitions of $w_0$, $w_{jd}$, and $w_{jj}$ in Section \ref{sec:theCode}. 
Let $Y^n$ denote the channel output when $X_{[k]}^n$ is transmitted.
Then the joint distribution of $(U_0^n,U_{[k]}^n,X_{[k]}^n,Y^n)$ is given by
\begin{equation*}
  p_\mathrm{code}(u_0^n,u_{[k]}^n,x_{[k]}^n,y^n)=
  p(u_0^n)p_\mathrm{code}(u_{[k]}^n|u_0^n)p(x_{[k]}^n|u_0^n,u_{[k]}^n)p(y^n|x_{[k]}^n), 
\end{equation*}
where 
\begin{equation*}
  p_\mathrm{code}(u_{[k]}^n|u_0^n)=
  \sum_{\mu_{[k]}}p(\mu_1|u_0^n)\dots p(\mu_k|u_0^n)
  p(u_{[k]}^n|u_0^n,\mu_{[k]})
\end{equation*}
and $p(\mu_j|u_0^n)$ and $p(u_{[k]}^n|u_0^n,\mu_{[k]})$ 
are calculated according to 
\begin{equation*}
  p(\mu_j|u_0^n)
	=\prod_{z_j\in [2^{nC_{jd}}]}p(\mu_j(z_j)|u_0^n),
\end{equation*}
and
\begin{equation*}
  p(u_{[k]}^n|u_0^n,\mu_{[k]})
	=\sum_{z_{[k]}}p(z_{[k]}|u_0^n,\mu_{[k]})
	\prod_{j=1}^k\mathbf{1}\{\mu_j(z_j)=u_j^n\}.
\end{equation*}
Define the distribution
$p_\mathrm{ind}(u_0^n,u_{[k]}^n,x_{[k]}^n,y^n)$ as
\begin{equation*}
  p_\mathrm{ind}(u_0^n,u_{[k]}^n,x_{[k]}^n,y^n)=
  p(u_0^n)p(x_{[k]}^n|u_0^n,u_{[k]}^n)p(y^n|x_{[k]}^n)
	\prod_{j=1}^k p(u_{j}^n|u_0^n),
\end{equation*}
which is the joint input-output distribution if independent codewords
are transmitted. We next mention some results regarding weakly typical
sets that are required for our error analysis.

For any $S\subseteq [k]$, 
let $A_\delta^{(n)}(U_0,U_S)$ denote the weakly typical set with respect to the
distribution $p(u_0,u_S)$, a marginal of $p(u_0,u_{[k]})$. 
In addition, for every $(u_0^n,u_S^n)\in A_\delta^{(n)}(U_0,U_S)$,
let $A_\delta^{(n)}(u_0^n,u_S^n)$ be the set of all $u_{S^c}^n$ such that 
\begin{equation*}
  (u_0^n,u_{[k]}^n)\in A_\delta^{(n)}(U_0,U_{[k]}).
\end{equation*}
Similarly, let $A_\epsilon^{(n)}(U_0,U_{[k]},X_{[k]},Y)$ be the weakly typical set with respect
to the distribution $p(u_0,u_{[k]},x_{[k]})p(y|x_{[k]})$, where $p(y|x_{[k]})$ is given by the channel
definition. For subsets $S,T\subseteq [k]$, 
define $A_\epsilon^{(n)}(U_0,U_S,X_T,Y)$ and $A_\epsilon^{(n)}(u_0^n,u_S^n,x_T^n,y^n)$ 
accordingly. If $(u_0^n,u_S^n,x_T^n,y^n)\in A_\epsilon^{(n)}(U_0,U_S,X_T,Y)$, we
have \cite[p. 523]{CoverThomas}
\begin{equation} \label{eq:typsetbound}
  \log|A_\epsilon^{(n)}(u_0^n,u_S^n,x_T^n,y^n)|
	\leq n\big(H(U_{S^c},X_{T^c}|U_0,U_S,X_T,Y)+2\epsilon\big).
\end{equation}
Finally, under fairly general conditions 
described in Appendix \ref{app:largeDev},\footnote{Distributions that 
satisfy these conditions include any distribution 
with finite support and the Gaussian distribution.} there exists
an increasing function $I:\mathbb{R}_{>0}\rightarrow\mathbb{R}_{>0}$ such
that if $(U_0^n,U_{[k]}^n,X_{[k]}^n,Y^n)$ consists of $n$ i.i.d.\ copies of 
$(U_0,U_{[k]},X_{[k]},Y)$ distributed according to $p(u_0,u_{[k]},x_{[k]},y)$, then 
\begin{equation} \label{eq:largeDev}
  \pr\Big\{(U_0^n,U_{[k]}^n,X_{[k]}^n,Y^n)\in 
	A_\epsilon^{(n)}(U_0,U_{[k]},X_{[k]},Y)\Big\}
  \geq 1-2^{-nI(\epsilon)}.
\end{equation}
Fix any such function $I$.

We next study the relationship between $p_\mathrm{code}$ and $p_\mathrm{ind}$. Our
first lemma provides an upper bound for $p_\mathrm{code}$ in terms
of $p_\mathrm{ind}$. 
\begin{lem} \label{lem:pcodepind}
For every nonempty $S\subseteq [k]$ and all $(u_0^n,u_S^n)$, 
\begin{equation*}
  \frac{1}{n}\log \frac{p_\mathrm{code}(u_S^n|u_0^n)}{p_\mathrm{ind}(u_S^n|u_0^n)}
  \leq nC_{Sd},
\end{equation*} 
where $C_{Sd}=\sum_{j\in S}C_{jd}$. 
\end{lem}
\begin{IEEEproof} Recall 
\begin{equation*}
  p_\mathrm{code}(u_S^n|u_0^n) =
  \sum_{\mu_{[k]}}p(u_S^n|u_0^n,\mu_{[k]})
	\prod_{j\in [k]}p(\mu_j|u_0^n).
\end{equation*}
To bound $p_\mathrm{code}(u_S^n|u_0^n)$, note that
\begin{equation*}
  p(u_S^n|u_0^n,\mu_{[k]})\leq
  \prod_{j\in S} \mathbf{1}\big\{\mu_j^{-1}(u_j^n)\neq \emptyset\big\},
\end{equation*}
where 
\begin{equation*}
  \mu_j^{-1}(u_j^n)=\big\{z_j\in [2^{nC_{jd}}]:
  \mu_j(z_j)=u_j^n\big\}.
\end{equation*}
Now for every $j\in S$,
\begin{align*}
  \sum_{\mu_j}p(\mu_j|u_0^n)\mathbf{1}\big\{\mu_j^{-1}(u_j^n)\neq \emptyset\big\}
  &= \pr\big\{\exists z_j:U_j^n(z_j)=u_j^n|U_0^n=u_0^n\big\}\\
  &\leq 2^{nC_{jd}}p(u_j^n|u_0^n).
\end{align*}
Thus
\begin{align*}
  p_\mathrm{code}(u_S^n|u_0^n)
  &\leq \sum_{\mu_S}\prod_{j\in S}p(\mu_j|u_0^n)
  \mathbf{1}\big\{\mu_j^{-1}(u_j^n)\neq\emptyset\big\}\\
  &=
  \prod_{j\in S} 
  \Big(\sum_{\mu_j}p(\mu_j|u_0^n)\mathbf{1}\big\{\mu_j^{-1}(u_j^n)\neq\emptyset\big\}\Big)\\
  &\leq 2^{n\sum_{j\in S}C_{jd}}p_\mathrm{ind}(u_S^n|u_0^n).
\end{align*}
\end{IEEEproof}
Our second lemma provides an upper bound for $p_\mathrm{ind}(u_S^n|u_0^n)$
when $(u_0^n,u_S^n)$ is typical. 
\begin{lem} \label{lem:pindub}
For all nonempty $S_d^c\subseteq S\subseteq [k]$ and 
$(u_0^n,u_S^n)\in A_\delta^{(n)}(U_0,U_S)$, 
\begin{equation*}
  \frac{1}{n}\log\frac{p_\mathrm{ind}(u_S^n|u_0^n)}{p(u_S^n|u_0^n)}\leq 
  -\sum_{j\in S\cap S_d}H(U_j|U_0)+H(U_{S\cap S_d}|U_0,U_{S_d^c})+2(|S\cap S_d|+1)\delta.
\end{equation*} 
\end{lem}
\begin{IEEEproof}
Recall that
\begin{equation*}
  p(u_{[k]}^n|u_0^n)
	=p(u_{S_d}^n|u_0^n,u_{S_d^c}^n)
	\prod_{j\in S_d^c}p(u_j^n|u_0^n).
\end{equation*} 
Thus for all $S\supseteq S_d^c$, we have
\begin{equation*}
  p(u_S^n|u_0^n)=p(u_{S\cap S_d}^n|u_0^n,u_{S_d^c}^n)
	\prod_{j\in S_d^c}p(u_j^n|u_0^n).
\end{equation*}
Therefore,
\begin{align*}
  \frac{p_\mathrm{ind}(u_S^n|u_0^n)}{p(u_S^n|u_0^n)}
  &=\frac{p_\mathrm{ind}(u_{S\cap S_d}^n|u_0^n)}{p(u_{S\cap S_d}^n|u_0^n,u_{S_d^c}^n)}\\
	&=\frac{\prod_{j\in S\cap S_d}p(u_j^n|u_0^n)}{p(u_{S\cap S_d}^n|u_0^n,u_{S_d^c}^n)}.
\end{align*}
The proof now follows from the definition of $A_\delta^{(n)}(U_0,U_S)$.
\end{IEEEproof}
Combining the previous two lemmas results in the next corollary, which we use 
in our error analysis. 
\begin{cor} \label{cor:pcodeub}
For every nonempty $S$ satisfying $S_d^c\subseteq S\subseteq [k]$
and all $(u_0^n,u_S^n)\in A_\delta^{(n)}(U_0,U_S)$, 
\begin{equation*}
  \frac{1}{n}\log\frac{p_\mathrm{code}(u_S^n|u_0^n)}{p(u_S^n|u_0^n)}
  \leq \zeta_{S\cap S_d}+2\big(|S\cap S_d|+1\big)\delta.
\end{equation*}
\end{cor}

Let $\mathcal{E}$ denote the event where either the output
of an encoder does not satisfy the corresponding cost constraint,
or the output of the decoder differs from the transmitted
$k$-tuple of messages; that is $(\hat{W}_j)_{j=1}^k\neq (w_j)_{j=1}^k$. 
Denote the former event with $\mathcal{E}_\mathrm{cost}$ and the 
latter event with $\mathcal{E}_\mathrm{dec}$. When $\mathcal{E}_\mathrm{dec}$ 
occurs, it is either the case that $(w_j)_{j=1}^k$ does not
satisfy (\ref{eq:dec}) (denote this event with $\mathcal{E}_\mathrm{typ}$), 
or that there is another $k$-tuple, $(\hat{W}_j)_{j=1}^k\neq (w_j)_{j=1}^k$, 
that also satisfies (\ref{eq:dec}).
If the latter event occurs, we either have 
$\hat{W}_0\neq w_0$ (denote event with $\mathcal{E}_{\emptyset,\emptyset}$), 
or $\hat{W}_0=w_0$. When $\hat{W}_0=w_0$, define the
subsets $S,T\subseteq [k]$ as
\begin{align*}
  S &= \big\{j:\hat{W}_{jd}\neq w_{jd}\big\}\\
  T &= \big\{j:\hat{W}_{jj}\neq w_{jj}\big\}.
\end{align*}
Now for every pair of subsets $S,T\subseteq [k]$ 
such that $S\cup T\neq \emptyset$, 
define $\mathcal{E}_{S,T}$ as the event where there exists a $(\hat{W}_j)_{j=1}^k$
that satisfies (\ref{eq:dec}), $\hat{W}_0=w_0$, 
$\hat{W}_{jd}\neq w_{jd}$ if and only if $j\in S$, and
$\hat{W}_{jj}\neq w_{jj}$ if and only if $j\in T$. Thus we may write
\begin{equation*}
  \mathcal{E}\subseteq \mathcal{E}_\mathrm{cost}\cup \mathcal{E}_\mathrm{typ}\cup 
  \bigcup_{S,T\subseteq [k]} \mathcal{E}_{S,T}.
\end{equation*}
The union over all $\mathcal{E}_{S,T}$ also contains the event
$\mathcal{E}_{\emptyset,\emptyset}$.
By the union bound, 
\begin{equation*}
  \pr(\mathcal{E})\leq 
	\pr(\mathcal{E}_\mathrm{cost})+
	\pr(\mathcal{E}_\mathrm{typ})
  + \sum_{S,T\subseteq [k]}\pr(\mathcal{E}_{S,T}).
\end{equation*}
Thus to find a set of achievable rates for our random code design, it
suffices to find conditions under which $\pr(\mathcal{E}_\mathrm{cost})$,
$\pr(\mathcal{E}_\mathrm{typ})$, and each
$\pr(\mathcal{E}_{S,T})$ go to zero as $n\rightarrow \infty$. 

We begin our analysis with the event $\mathcal{E}_\mathrm{cost}$.
For $j\in [k]$, let $\mathcal{E}_\mathrm{cost}^j$ denote the event
where the codeword $X_j^n(w_{jj}|U_0^n(w_0),U_j^n(w_{jd},Z_j))$
does not satisfy the cost constraint of encoder $j$. We have
\begin{align*}
  \pr\big(\mathcal{E}^j_\mathrm{cost}\big)
  &= \pr\left\{\frac{1}{n}\sum_{t=1}^n 
	b_j\Big(X_{jt}\big(w_{jj}|U_0^n(w_0),U_j^n(w_{jd},Z_j)\big)\Big)> B_j\right\}\\
	&= \sum_{z_j}\pr\{Z_j=z_j\}
	\pr\left\{\frac{1}{n}\sum_{t=1}^n 
	b_j\Big(X_{jt}\big(w_{jj}|U_0^n(w_0),U_j^n(w_{jd},z_j)\big)\Big)> B_j\right\}.
\end{align*}
Since for all $z_j$, by the AEP,
\begin{equation*}
  \pr\left\{\frac{1}{n}\sum_{t=1}^n 
	b_j\Big(X_{jt}\big(w_{jj}|U_0^n(w_0),U_j^n(w_{jd},z_j)\big)\Big)> B_j\right\}
	\rightarrow 0
\end{equation*}
as $n\rightarrow\infty$, it follows that 
$\pr\big(\mathcal{E}^j_\mathrm{cost}\big)\rightarrow 0$. 
Applying the union bound now implies
\begin{equation*}
  \pr\big(\mathcal{E}_\mathrm{cost}\big)
	\leq \sum_{j\in [k]}\pr\big(\mathcal{E}^j_\mathrm{cost}\big)
	\rightarrow 0. 
\end{equation*}

We next consider the event $\mathcal{E}_\mathrm{typ}$. 
Define $\mathcal{E}_\mathrm{enc}$ as the event where 
\begin{equation*}
  \big(U_0^n,U_{[k]}^n\big)\notin A_\delta^{(n)}(U_0,U_{[k]})
\end{equation*}
and note that $\mathcal{E}_\mathrm{typ}$ is the event where
\begin{equation*}
  \big(U_0^n,U_{[k]}^n,X_{[k]}^n,Y^n\big)
  \notin A_\epsilon^{(n)}(U_0,U_{[k]},X_{[k]},Y).
\end{equation*}
The event $\mathcal{E}_\mathrm{enc}$ occurs if and only if 
$\mathcal{A}(U_0^n,U_{[k]}^n(.))$ (defined in Section \ref{sec:theCode})
is empty. Thus
\begin{equation*}
  \pr(\mathcal{E}_\mathrm{enc})
  =\pr\big\{\mathcal{A}(U_0^n,U_{[k]}^n(.))=\emptyset\big\}.
\end{equation*}
If $S_d=\emptyset$, $\pr(\mathcal{E}_\mathrm{enc})$  
goes to zero by the AEP since in this case 
$p_\mathrm{code}(u_{[k]}^n|u_0^n)=p(u_{[k]}^n|u_0^n)$. 
Otherwise, recall that for every
nonempty $S\subseteq S_d$, $\zeta_S$ is defined as
\begin{equation*}
  \zeta_S = \sum_{j\in S}C_{jd}-
  \sum_{j\in S}H(U_j|U_0)+H(U_S|U_0,U_{S_d^c}).
\end{equation*}
From the multivariate covering lemma (Appendix \ref{app:MCL}),
it follows that $\pr(\mathcal{E}_\mathrm{enc})\rightarrow 0$
if for all nonempty $S\subseteq S_d$,
\begin{equation} \label{eq:zetalb}
  \zeta_S > (8|S_d|-2|S|+10)\delta.
\end{equation} 

Next we find an upper bound for 
$\pr(\mathcal{E}_\mathrm{typ}\setminus\mathcal{E}_\mathrm{enc})$. Let
$B^{(n)}$ be the set of all $(u_0^n,u_{[k]}^n,x_{[k]}^n,y^n)$ such that 
$(u_0^n,u_{[k]}^n)\in A_\delta^{(n)}$
but $(u_0^n,u_{[k]}^n,x_{[k]}^n,y^n)\notin A_\epsilon^{(n)}$. Then 
\begin{align*}
  \pr(\mathcal{E}_\mathrm{typ}\setminus\mathcal{E}_\mathrm{enc})
  &= \sum_{B^{(n)}}p(u_0^n)p_\mathrm{code}(u_{[k]}^n|u_0^n)
  p(x_{[k]}^n|u_0^n,u_{[k]}^n)p(y^n|x_{[k]}^n)\\
  &\overset{(a)}{\leq} 2^{n(\zeta_{S_d}+2(|S_d|+1)\delta)}\sum_{B^{(n)}}
  p(u_0^n,u_{[k]}^n,x_{[k]}^n,y^n)\\
  &\overset{(b)}{\leq} 2^{n(\zeta_{S_d}+2(|S_d|+1)\delta)}\pr\big\{(A_\epsilon^{(n)})^c\big\}
  \overset{(c)}{\leq} 2^{n(\zeta_{S_d}+2(|S_d|+1)\delta-I(\epsilon))},
\end{align*}
where (a) follows from Corollary \ref{cor:pcodeub}, (b) holds since $B^{(n)}\subseteq 
(A_\epsilon^{(n)})^c$, and (c) follows from the definition of $I(\epsilon)$ 
given by (\ref{eq:largeDev}).
Thus $\pr(\mathcal{E}_\mathrm{typ}\setminus\mathcal{E}_\mathrm{enc})\rightarrow 0$ if 
\begin{equation} \label{eq:zetaub}
  \zeta_{S_d}<I(\epsilon)-2(|S_d|+1)\delta.
\end{equation}
Therefore, if (\ref{eq:zetalb}) and (\ref{eq:zetaub}) both 
hold, then $\pr(\mathcal{E}_\mathrm{typ})\rightarrow 0$ since 
\begin{equation*}
  \pr(\mathcal{E}_\mathrm{typ})\leq \pr(\mathcal{E}_\mathrm{enc}\cup \mathcal{E}_\mathrm{typ})
  =\pr(\mathcal{E}_\mathrm{enc})+\pr(\mathcal{E}_\mathrm{typ}\setminus\mathcal{E}_\mathrm{enc}).
\end{equation*}

We next study $\mathcal{E}_{\emptyset,\emptyset}$, which is the event where there exists
a $k$-tuple $(\hat{W}_j)_j$ that satisfies (\ref{eq:dec}) but $\hat{W}_0\neq w_0$.
If this event occurs, then $(\hat{U}_0^n,\hat{U}_{[k]}^n,\hat{X}_{[k]}^n)$
and $Y^n$ are independent. By the union bound,
\begin{align*}
  \pr(\mathcal{E}_{\emptyset,\emptyset}) &\leq 
  2^{n\sum_{j=1}^kR_j}\sum_{A_\epsilon^{(n)}}
  p_\mathrm{code}(u_0^n,u_{[k]}^n,x_{[k]}^n)p_\mathrm{code}(y^n).
\end{align*}  
We rewrite the sum in the above inequality as
\begin{equation*}
  \sum_{A_\epsilon^{(n)}(Y)}p_\mathrm{code}(y^n)
  \sum_{A_\epsilon^{(n)}(y^n)}p_\mathrm{code}(u_0^n,u_{[k]}^n,x_{[k]}^n),
\end{equation*}
Using Corollary \ref{cor:pcodeub}, we upper bound the inner sum by
\begin{align*}
  \MoveEqLeft
  \sum_{A_\epsilon^{(n)}(y^n)}2^{n(\zeta_{S_d}+2(|S_d|+1)\delta)}
  p(u_0^n,u_{[k]}^n,x_{[k]}^n)\\
  &\overset{(*)}{\leq} 2^{n(H(U_0,U_{[k]},X_{[k]}|Y)+2\epsilon)}
	2^{n(\zeta_{|S_d|}+2(|S_d|+1)\delta)}
  2^{-n(H(U_0,U_{[k]},X_{[k]})+\epsilon)},
\end{align*}
where $(*)$ follows from (\ref{eq:typsetbound}). 
This implies $\pr(\mathcal{E}_{\emptyset,\emptyset})\rightarrow 0$ if
\begin{equation*}
  \sum_{j=1}^k R_j < I(X_{[k]};Y)-\zeta_{S_d}-2(|S_d|+1)\delta-3\epsilon.
\end{equation*}

Next, let $S,T\subseteq [k]$ be sets such that $S\cup T\neq \emptyset$
and consider the event $\mathcal{E}_{S,T}$.
Recall that this is the event where there exists a $k$-tuple $(\hat{W}_j)_j$
that satisfies (\ref{eq:dec})
and $\hat{W}_0=w_0$, $\hat{W}_{jd}\neq w_{jd}$ if and only if $j\in S$, 
and $\hat{W}_{jj}\neq w_{jj}$ if and only if $j\in T$.
For every $A\subseteq S$ and $B\subseteq S^c$, 
let $\mathcal{E}_{S,T}^{A,B}\subseteq\mathcal{E}_{S,T}$ be the event where 
there exists a $k$-tuple $(\hat{W}_j)_j$ that satisfies
\begin{align} \label{eq:ESTABdef}
  \MoveEqLeft
  \Big(U_0^n(w_0),\big(U_j^n(\hat{W}_{jd},\hat{Z}_j|U_0^n)\big)_{j\in A},
	\big(U_j^n(w_{jd},\hat{Z}_j|U_0^n)\big)_{j\in B},\notag\\
  &\big(X_j^n(\hat{W}_{jj}|U_0^n,\hat{U}_j^n)\big)_{j\in A\cup (B\cap T)},
	\big(X_j^n(w_{jj}|U_0^n,\hat{U}_j^n)\big)_{j\in B\setminus T},
  Y^n\Big)\in A_\epsilon^{(n)}
\end{align}
and $\hat{W}_0=w_0$, $\hat{W}_{jd}\neq w_{jd}$ if and only if $j\in S$, 
and $\hat{W}_{jj}\neq w_{jj}$ if and only if $j\in T$.
If $\mathcal{E}_{S,T}$ occurs, then so does $\mathcal{E}_{S,T}^{A,B}$ for
every $A\subseteq S$ and $B\subseteq S^c$. Thus
\begin{equation*}
  \mathcal{E}_{S,T}\subseteq\bigcap_{A,B}
	\mathcal{E}_{S,T}^{A,B}.
\end{equation*}
This implies
\begin{equation} \label{eq:ESTminAB}
  \pr(\mathcal{E}_{S,T})\leq \min_{A,B}
	\pr\big(\mathcal{E}_{S,T}^{A,B}\big).
\end{equation}
Therefore, to bound $\pr(\mathcal{E}_{S,T})$, we find an upper bound
on $\pr(\mathcal{E}_{S,T}^{A,B})$ for any $A\subseteq S$ and $B\subseteq S^c$ such
that $A\cup (B\cap T)\neq \emptyset$. This is the key difference between 
our error analysis here and the error analysis for the 2-user MAC with
transmitter cooperation presented in \cite{Noorzad2}. For independent 
distributions, using the constraint that subsets of typical codewords are 
also typical does not lead to a larger region; the same may not be true 
when dealing with dependent distributions. That being said, to include
all independent random variables in our error analysis, instead of
calculating the minimum in (\ref{eq:ESTminAB}) over all 
$A\subseteq S$ and $B\subseteq S^c$,
we limit ourselves to subsets $A$ and $B$ that satisfy
\begin{align*}
  S\cap S_d^c &\subseteq A\subseteq S\\
  S^c\cap S_d^c &\subseteq B\subseteq S^c,
\end{align*}
since all the random vectors $(U_j^n)_{j\in S_d^c}$ are independent given $U_0^n$. 
Choose any such $A$ and $B$. Note that for every $j\in A\cup (B\cap T)$, 
either $\hat{W}_{jd}\neq w_{jd}$ or $\hat{W}_{jj}\neq w_{jj}$. 
In addition, in (\ref{eq:ESTABdef}), 
\begin{align*} 
  \MoveEqLeft
  \Big(\big(U_j^n(\hat{W}_{jd},\hat{Z}_j|U_0^n)\big)_{j\in A},
	\big(U_j^n(w_{jd},\hat{Z}_j|U_0^n)\big)_{j\in B},\notag\\
  &\big(X_j^n(\hat{W}_{jj}|U_0^n,U_j^n)\big)_{j\in A\cup (B\cap T)},
	\big(X_j^n(w_{jj}|U_0^n,U_j^n)\big)_{j\in B\setminus T}\Big)
\end{align*}
is independent of $Y^n$ given
\begin{equation*}
\Big(U_0^n(w_0),\big(U_j^n(w_{jd},.|U_0^n)\big)_{j\in S^c},
  \big(X_j^n(w_{jj}|U_0^n,U_j^n(.))\big)_{j\in S^c\setminus T}\Big).
\end{equation*}
Therefore, by the union bound, $\pr(\mathcal{E}_{S,T}^{A,B})$ is bounded from above
by
\begin{align} \label{eq:Est}
  \MoveEqLeft
  2^{n\big(\sum_{j\in A}R_{jd}+\sum_{j\in A\cup (B\cap T)}R_{jj}\big)}\notag\\
  &\times\sum_{A_\epsilon^{(n)}}p(x_{A\cup (B\cap T)}^n|u_0^n,u_{A\cup (B\cap T)}^n)
  \notag\\
  &\phantom{\times\sum_{A_\epsilon^{(n)}}}
  \times\smashoperator[r]{\sum_{\mu_{A\cup S^c},\chi_{S^c\setminus T}}}
  p(u_0^n,\mu_{S^c},\chi_{S^c\setminus T},y^n)p(\mu_A|u_0^n)
  p(u_{A\cup B}^n,x_{B\setminus T}^n|u_0^n,\mu_{A\cup S^c},\chi_{S^c\setminus T}),
\end{align}
where the inner sum is over all mappings 
$\mu_j:[2^{nC_{jd}}]\rightarrow\mathcal{U}_j^n$ 
for $j\in A\cup S^c$ and $\chi_j:[2^{nC_{jd}}]\rightarrow\mathcal{X}_j^n$
for $j\in S^c\setminus T$. The distribution 
$p(u_0^n,\mu_{S^c},\chi_{S^c\setminus T},y^n)$
is a marginal of $p(u_0^n,\mu_{[k]},\chi_{[k]},y^n)$, which is defined as
\begin{equation*}
  p(u_0^n,\mu_{[k]},\chi_{[k]},y^n)
	=p(u_0^n,\mu_{[k]})p(\chi_{[k]}|u_0^n,\mu_{[k]})
	p(y^n|u_0^n,\mu_{[k]},\chi_{[k]}),
\end{equation*}
where 
\begin{align*}
  p(\chi_{[k]}|u_0^n,\mu_{[k]})
	&= \prod_{j\in [k]}p(\chi_j|u_0^n,\mu_j)\\
	&= \prod_{j\in [k]}\prod_{z_j\in [2^{nC_{jd}}]}
	p(\chi_j(z_j)|u_0^n,\mu_j(z_j)),
\end{align*}
and
\begin{equation*}
  p(y^n|u_0^n,\mu_{[k]},\chi_{[k]})
	=\sum_{z_{[k]}}p(z_{[k]}|u_0^n,\mu_{[k]})
	p(y^n|\chi_{[k]}(z_{[k]})).
\end{equation*}
We have
\begin{align} \label{eq:ST}
  \MoveEqLeft
  p(u_{A\cup B}^n,x_{B\setminus T}^n|u_0^n,\mu_{A\cup S^c},\chi_{S^c\setminus T})\notag\\
  &\leq \mathbf{1}\Big\{\exists (z_j)_{j\in B}\in\prod_{j\in B}[2^{nC_{jd}}]:
  (\forall j\in B:\mu_j(z_j)=u_j^n)\wedge 
  (\forall j\in B\setminus T:\chi_j(z_j)=x_j^n)\Big\}\notag\\
  &\phantom{\leq}\times\mathbf{1}\Big\{\exists (z_j)_{j\in A}\in\prod_{j\in A}[2^{nC_{jd}}]:
  (\forall j\in A:\mu_j(z_j)=u_j^n)\Big\}.
\end{align}
We can thus upper bound the inner sum in (\ref{eq:Est}) as a 
product of the sums 
\begin{align*}
  \MoveEqLeft
  \sum_{\mu_{S^c},\chi_{S^c\setminus T}}
  p(u_0^n,\mu_{S^c},\chi_{S^c\setminus T},y^n)\\
  &\times\mathbf{1}\Big\{\exists (z_j)_{j\in B}\in\prod_{j\in B}[2^{nC_{jd}}]:
  (\forall j\in B:\mu_j(z_j)=u_j^n)\wedge 
  (\forall j\in B\setminus T:\chi_j(z_j)=x_j^n)\Big\} 
\end{align*}
and 
\begin{equation*}
  \sum_{\mu_A}p(\mu_A|u_0^n) 
  \mathbf{1}\Big\{\exists (z_j)_{j\in A}\in\prod_{j\in A}[2^{nC_{jd}}]:
  \forall j\in A, \mu_j(z_j)=u_j^n\Big\}.
\end{equation*}
We first find an upper bound for the first sum.
Define the distribution $\tilde{p}(u_0^n,u_{[k]}^n,x_{[k]}^n,y^n)$ as
\begin{equation*}
  \tilde{p}(u_0^n,u_{[k]}^n,x_{[k]}^n,y^n)
  = \sum_{\mu_{[k]},\chi_{[k]}}p(u_0^n,\mu_{[k]},\chi_{[k]},y^n)\prod_{j=1}^k
  \mathbf{1}\big\{\mu_j(1)=u_j^n,\chi_j(1)=x_j^n\big\}.
\end{equation*}
The following argument demonstrates that 
$\tilde{p}(u_0^n,u_{[k]}^n,x_{[k]}^n)=p_\mathrm{ind}(u_0^n,u_{[k]}^n,x_{[k]}^n)$, 
\begin{align} \label{eq:ptildepind}
  \tilde{p}(u_0^n,u_{[k]}^n,x_{[k]}^n) 
  &= \sum_{y^n}\tilde{p}(u_0^n,u_{[k]}^n,x_{[k]}^n,y^n)\notag\\
  &= \sum_{\mu_{[k]},\chi_{[k]}} p(u_0^n,\mu_{[k]},\chi_{[k]})\prod_{j=1}^k
  \mathbf{1}\big\{\mu_j(1)=u_j^n,\chi_j(1)=x_j^n\big\}\notag\\
  &= p(u_0^n)\prod_{j=1}^k \sum_{\mu_j,\chi_j}p(\mu_j,\chi_j|u_0^n)
  \mathbf{1}\big\{\mu_j(1)=u_j^n,\chi_j(1)=x_j^n\big\}\notag\\
  &=p_\mathrm{ind}(u_0^n,u_{[k]}^n,x_{[k]}^n).
\end{align}
For every $\mathbf{z}=(z_j)_{j\in B}$, where $z_j\in [2^{nC_{jd}}]$ 
for all $j\in B$, 
let $E_\mathbf{z}$ denote the event where for all $j\in B$, 
$U_j^n(w_{jd},z_j|U_0^n)=u_j^n$, and for all $j\in B\setminus T$, 
$X_j^n(w_{jj}|U_0^n,U_j^n)=x_j^n$. Then
\begin{align} \label{eq:T}
  \MoveEqLeft
  \sum_{\mu_{S^c},\chi_{S^c\setminus T}}
  p(u_0^n,\mu_{S^c},\chi_{S^c\setminus T},y^n)\notag\\
  &\times\mathbf{1}\Big\{\exists \mathbf{z}\in\prod_{j\in B}[2^{nC_{jd}}]:
  (\forall j\in B:\mu_j(z_j)=u_j^n)\wedge 
  (\forall j\in B\setminus T:\chi_j(z_j)=x_j^n)\Big\}\notag \\
  &=\pr\Big(\{U_0^n=u_0^n, Y^n=y^n\} \cap \bigcup_\mathbf{z}E_\mathbf{z}\Big)\notag\\
  &=\pr\Big(\bigcup_\mathbf{z}\big(\{U_0^n=u_0^n, Y^n=y^n\} \cap E_\mathbf{z}\big)\Big)\notag\\
  &\overset{(a)}{\leq} 2^{nC_{Bd}}\pr\big(\{U_0^n=u_0^n, Y^n=y^n\}\cap E_{\mathbf{z}=\mathbf{1}}\big)\notag\\
  &= 2^{nC_{Bd}}\tilde{p}(u_0^n,u_B^n,x_{B\setminus T}^n,y^n)\notag\\
  &\overset{(b)}{=} 2^{nC_{Bd}}p(u_0^n)p_\mathrm{ind}(u_B^n,x_{B\setminus T}^n|u_0^n)
	\tilde{p}(y^n|u_0^n,u_B^n,x_{B\setminus T}^n),
\end{align}
where $(a)$ follows by the union bound and $(b)$ follows by (\ref{eq:ptildepind}).
Using a similar argument we can show
\begin{equation} \label{eq:S}
  \sum_{\mu_A}p(\mu_A|u_0^n) 
  \mathbf{1}\Big\{\exists \mathbf{z}\in\prod_{j\in A}[2^{nC_{jd}}]:
  \forall j\in A, \mu_j(z_j)=u_j^n\Big\}
  \leq 2^{nC_{Ad}}p_\mathrm{ind}(u_A^n|u_0^n).
\end{equation}
Thus by (\ref{eq:ST}), (\ref{eq:T}), and (\ref{eq:S}), the expression
\begin{align*}
  \MoveEqLeft 
  2^{n\big(\sum_{j\in A}R_{jd}+\sum_{j\in A\cup (B\cap T)}R_{jj}+C_{Ad}+C_{Bd}\big)}\\
  &\times\sum_{A_\epsilon^{(n)}} p(u_0^n)p_\mathrm{ind}(u_{A\cup B}^n|u_0^n)
  p(x_{A\cup B}^n|u_0^n,u_{A\cup B}^n)
  \tilde{p}(y^n|u_0^n,u_B^n,x_{B\setminus T}^n)
\end{align*}
is an upper bound for (\ref{eq:Est}). Applying Lemma \ref{lem:pindub}
to $p_\mathrm{ind}(u_{A\cup B}^n|u_0^n)$ and dropping the epsilon term, 
this expression can be further bounded from above by
\begin{align*}
  \MoveEqLeft
	2^{n\big(\sum_{j\in A}R_{jd}+\sum_{j\in A\cup (B\cap T)}R_{jj}+\zeta_{(A\cup B)\cap S_d}\big)}\\
	&\phantom{=}\times
  \smashoperator[r]{\sum_{A_\epsilon^{(n)}(U_0,U_B,X_{B\setminus T},Y)}}
  p(u_0^n,u_B^n,x_{B\setminus T}^n)\tilde{p}(y^n|u_0^n,u_B^n,x_{B\setminus T}^n)\\
  &\phantom{=}\times
  \smashoperator[r]{\sum_{A_\epsilon^{(n)}(u_0^n,u_B^n,x_{B\setminus T}^n,y^n)}}
  p(u_A^n|u_0^n,u_B^n)p(x_{A\cup (B\cap T)}^n|u_0^n,u_{A\cup (B\cap T)}^n)
\end{align*}
Using (\ref{eq:typsetbound}),
we can further upper bound the logarithm of this expression by
\begin{align*}
  \MoveEqLeft
  n\bigg[\sum_{j\in A}R_{jd}+\sum_{j\in A\cup (B\cap T)}R_{jj}+\zeta_{(A\cup B)\cap S_d}\bigg]\\
  &+\log
	\smashoperator[r]{\sum_{A_\epsilon^{(n)}(U_0,U_B,X_{B\setminus T},Y)}}
	p(u_0^n,u_B^n,x_{B\setminus T}^n)\tilde{p}(y^n|u_0^n,u_B^n,x_{B\setminus T}^n)\\
  &-nH(U_A|U_0,U_B)-nH(X_{A\cup (B\cap T)}|U_0,U_{A\cup (B\cap T)})\\
	&+nH(U_A,X_{A\cup (B\cap T)}|U_0,U_B,X_{B\setminus T},Y)
\end{align*}
Hence $\pr(\mathcal{E}_{S,T}^{A,B})\rightarrow 0$ if 
\begin{align*}
  \MoveEqLeft
  \sum_{j\in A}R_{jd}+\sum_{j\in A\cup (B\cap T)}R_{jj} \\
  &< -\zeta_{(A\cup B)\cap S_d}+H(U_A|U_0,U_B)+H(X_{A\cup (B\cap T)}|U_0,U_{A\cup (B\cap T)})\\
  &\phantom{<} -H(U_A,X_{A\cup (B\cap T)}|U_0,U_B,X_{B\setminus T},Y)\\
  &= I(U_A,X_{A\cup (B\cap T)};Y|U_0,U_B,X_{B\setminus T})-\zeta_{(A\cup B)\cap S_d},
\end{align*}
where the last equality follows from the fact that 
\begin{align*} 
  H(U_A|U_0,U_B) &= H(U_A|U_0,U_B,X_{B\setminus T})
	+I(U_A;X_{B\setminus T}|U_0,U_B)\\
	&=H(U_A|U_0,U_B,X_{B\setminus T})
\end{align*}
and
\begin{align*} 
  H(X_{A\cup(B\cap T)}|U_0,U_{A\cup B}) &= H(X_{A\cup(B\cap T)}|U_0,U_{A\cup B},X_{B\setminus T})
	+I(X_{A\cup(B\cap T)};X_{B\setminus T}|U_0,U_{A\cup B})\\
	&=H(X_{A\cup(B\cap T)}|U_0,U_{A\cup B},X_{B\setminus T}).
\end{align*}
Thus $\pr(\mathcal{E}_{S,T})\rightarrow 0$ if for some
$S\cap S_d^c\subseteq A\subseteq S$ and 
$S^c\cap S_d^c\subseteq B\subseteq S^c$ 
such that $A\cup (B\cap T)\neq\emptyset$,
\begin{align} \label{eq:main}
	\MoveEqLeft
  \sum_{j\in A}R_{jd}+\sum_{j\in A\cup (B\cap T)}R_{jj}\notag\\
  &<I(U_A,X_{A\cup (B\cap T)};Y|U_0,U_B,X_{B\setminus T})
	-\zeta_{(A\cup B)\cap S_d}
\end{align}

The bounds we obtain above are in terms of $(R_{jd})_{j=1}^k$ and $(R_{jj})_{j=1}^k$.
To convert these to bounds in terms of $(R_j)_{j=1}^k$, recall that 
$R_{j0}=\min\{C_{j0},R_j\}$, $R_{jj}=(R_j-C_\mathrm{in}^j)^+$, and
\begin{align*}
  R_{jd} &= R_j-R_{j0}-R_{jj}\\
  &= R_j-\min\{C_{j0},R_j\}-R_{jj}=\max\{R_j-C_{j0},0\}-(R_j-C_\mathrm{in}^j)^+\\
  &= (R_j-C_{j0})^+-(R_j-C_\mathrm{in}^j)^+.
\end{align*}
Thus (\ref{eq:main}) can be written as
\begin{align*} 
  \MoveEqLeft
  \sum_{j\in A}(R_j-C_{j0})^++\sum_{j\in B\cap T}(R_j-C_\mathrm{in}^j)^+\notag\\
  &< I(U_A,X_{A\cup (B\cap T)};Y|U_0,U_B,X_{B\setminus T})
	-\zeta_{(A\cup B)\cap S_d}
\end{align*}

\subsection{Theorem \ref{thm:sumCapacity} (Sum-capacity gain)} \label{subsec:sumCapacity}

Fix any unit vector $\mathbf{v}\in \mathbb{R}^k_{>0}$, rate vector
$\mathbf{C}_\mathrm{in}\in \mathbb{R}^k_{>0}$, and $\mathbf{B}\in\mathbb{R}^k_{\geq 0}$. 
For every $h\geq 0$, define
$\mathbf{C}_\mathrm{out}(h)=h\mathbf{v}$.  
In the achievable region defined in Section \ref{sec:model},
let $\mathcal{U}_0=\{0,1\}$, 
and for every $j\in [k]$, let $\mathcal{U}_j=\mathcal{X}_j$.
Set $C_{j0}=0$ and $C_{jd}=C_\mathrm{out}^j(h)$ for every $j\in [k]$.  
For $h>0$, let $\mathcal{P}(h)$ be the set of all distributions of the form
\begin{equation*}
	p(u_0,u_{[k]})\cdot
	\prod_{j\in [k]}p(x_j|u_0,u_j)
\end{equation*} 
that satisfy dependence constraints
\begin{equation*}
  \sum_{j\in S} C_\mathrm{out}^j(h)-\sum_{j\in S} H(U_j|U_0)
  +H(U_S|U_0)>0
	\qquad\forall\: \emptyset\subsetneq S\subseteq [k],
\end{equation*}
and cost constraints
\begin{equation*}
  \mathbb{E}\big[b_j(X_j)\big]\leq B_j
	\qquad \forall j\in [k]. 
\end{equation*}

Using Lemma \ref{lem:plus} (see end of this section), we see that 
every rate vector $(R_j)_{j\in[k]}$ that 
for some distribution $p\in\mathcal{P}(h)$ and 
every pair of subsets $S,T\subseteq [k]$ satisfies
\begin{equation} \label{eq:simpIB}
  \sum_{j\in S\cup T}R_j
	< I(X_{S\cup T};Y|U_0,U_{S^c},X_{S^c\cap T^c})
	+\sum_{j\in T\setminus S}C_\mathrm{in}^j
	-\zeta_{[k]}
\end{equation}
and 
\begin{equation*}
  \sum_{j\in [k]}R_j
	< I(X_{[k]};Y)-\zeta_{[k]},
\end{equation*}
is achievable. This follows from setting $A=S$ and $B=S^c$ 
for every $S,T\subseteq [k]$ in (\ref{eq:thmIB}).
To obtain a lower bound on sum-capacity, we evaluate this 
region for a specific distribution in 
$\mathcal{P}(h)$.

Since our MAC is in $\mathcal{C}^*(\mathcal{X}_{[k]},\mathcal{Y})$, 
there exists a distribution 
$p_a\in\mathcal{P}_\mathrm{ind}(\mathcal{X}_{[k]})$
that satisfies
\begin{equation*}
  I_a(X_{[k]};Y) =
  \max_{p\in\mathcal{P}_\mathrm{ind}(\mathcal{X}_{[k]})}I(X_{[k]};Y),
\end{equation*}
and a distribution $p_b\in\mathcal{P}(\mathcal{X}_{[k]})$ that 
satisfies
\begin{equation*}
  \mathbb{E}_b\Big[D\big(p(y|X_{[k]})\|p_a(y)\big)\Big]
	>\mathbb{E}_a\Big[D\big(p(y|X_{[k]})\|p_a(y)\big)\Big],
\end{equation*}
and whose support is contained in the support of $p_a$. 
Here we also assume that for all $j\in [k]$, 
\begin{equation*} 
  I_a(X_j;Y|X_{[k]\setminus\{j\}})>0.
\end{equation*}
At the end of the proof, we show that in
the case where this property does not hold, the same 
result follows by considering a MAC with a smaller number of 
users. 

Choose $\mu\in (0,1)$ such that for every nonempty $S\subseteq [k]$,
\begin{equation} \label{eq:muDef}
  \mu I_a(X_S;Y|X_{S^c})
	<\sum_{j\in S}C_\mathrm{in}^j.
\end{equation} 
For every $\lambda\in [0,1]$, define the distribution
$p_\lambda(u_0,u_{[k]},x_{[k]})$ as
\begin{equation*}
p_\lambda(u_0,u_{[k]},x_{[k]})=
p_\lambda(u_0)p_\lambda(u_{[k]})
p_\lambda(x_{[k]}|u_0,u_{[k]}),
\end{equation*}
where 
\begin{equation*}
  p_\lambda(u_0)
  =\begin{cases}
  \mu &\text{if }u_0=1\\
  1-\mu &\text{if }u_0=0,
  \end{cases}
\end{equation*}	
and for every $u_{[k]}\in\mathcal{U}_{[k]}$ 
(recall $\mathcal{U}_{[k]}=\mathcal{X}_{[k]}$),  
\begin{equation*}
  p_\lambda(u_{[k]})
  = (1-\lambda)p_a(u_{[k]})
  + \lambda p_b(u_{[k]}).
\end{equation*}
Finally, for every $(u_0,u_{[k]},x_{[k]})$,  
\begin{equation*}
  p_\lambda(x_{[k]}|u_0,u_{[k]})
  =\prod_{j=1}^k p_\lambda(x_j|u_0,u_j),
\end{equation*}
where for all $j\in [k]$,
\begin{equation*}
  p_\lambda(x_j|u_0,u_j)=
  \begin{cases}
  \mathbf{1}\{x_j=u_j\} &\text{if }u_0=1\\
  p_a(x_j) &\text{if }u_0=0.
  \end{cases}
\end{equation*} 
Note that $p_\lambda(u_0)$ and $p_\lambda(x_{[k]}|u_0,u_{[k]})$
do not depend on $\lambda$. In addition, since $p_a$ and $p_b$
satisfy the cost constraints and
\begin{equation*}
  p_\lambda(x_{[k]})=(1-\lambda)p_a(x_{[k]})
	+\lambda p_b(x_{[k]}),
\end{equation*}
for all $\lambda\in (0,1)$, $p_\lambda$ satisfies the 
cost constraints as well.

We next find a function $\lambda^*(h)$ so that 
\begin{equation*}
  p_{\lambda^*(h)}(u_0,u_{[k]},x_{[k]})\in
  \mathcal{P}(h)
\end{equation*}
for sufficiently small $h$. 
Fix $\epsilon>0$, and consider the equation
\begin{equation} \label{eq:lambdadef}
  h\sum_{j\in [k]}v_j
  =\sum_{j\in [k]}H_\lambda(U_j)-H_\lambda(U_{[k]})
  +\epsilon\lambda\sum_{j\in [k]}v_j.
\end{equation}
By Lemma \ref{lem:derivative} (see end of this section),
\begin{equation*}
  \frac{dh}
  {d\lambda}\Big|_{\lambda =0^+}=\epsilon> 0.
\end{equation*}
Thus the inverse function theorem implies that there 
exists a function $\lambda=\lambda^*(h)$
defined on $[0,h_0)$ for some $h_0>0$ that 
satisfies (\ref{eq:lambdadef}), and
\begin{equation} \label{eq:dlambda}
  \frac{d\lambda^*}{dh}\Big|_{h=0^+}=\frac{1}{\epsilon}.
\end{equation}
For every nonempty $S\subseteq [k]$, define the function 
$\zeta_S:[0,h_0)\rightarrow \mathbb{R}$ as 
\begin{equation} \label{eq:zetaSh}
  \zeta_S(h)
	= \sum_{j\in S} C^j_\mathrm{out}(h)
	-\sum_{j\in S} H_{\lambda^*}(U_j)
	+ H_{\lambda^*}(U_{S}),
\end{equation}
If we calculate the derivative of $\zeta_S$ at 
$h=0$, by Lemma \ref{lem:derivative}, we get
\begin{equation*} 
  \frac{d\zeta_S}{dh}\Big|_{h=0^+}
  =\sum_{j\in S}v_j>0.
\end{equation*}
This implies that there exists $0<h_1\leq h_0$ such that for every 
$0<h<h_1$ and all nonempty $S\subseteq [k]$,
\begin{equation*}
  \zeta_S(h)>0.
\end{equation*}
Therefore, for all sufficiently small $h$, 
$p_{\lambda^*(h)}(u_0,u_{[k]},x_{[k]})$ is in $\mathcal{P}(h)$. 

We next find a lower bound for the achievable sum-rate using the distribution 
$p_{\lambda^*}(u_0,u_{[k]},x_{[k]})$ for small $h$.
For every $S,T\subseteq [k]$, define the function 
$f_{S,T}:[0,h_1)\rightarrow\mathbb{R}$ as 
\begin{equation*}
  f_{S,T}(h)=I_{\lambda^*}(X_{S\cup T};Y|U_0,U_{S^c},X_{S^c\cap T^c})
  +\sum_{j\in T\setminus S}C_\mathrm{in}^j-\zeta_{[k]}(h).
\end{equation*}
In the above equation, expanding the mutual information term 
with respect to $U_0$ gives
\begin{align*} 
  \MoveEqLeft
  I_{\lambda^*}(X_{S\cup T};Y|U_0,U_{S^c},X_{S^c\cap T^c}) \notag\\
	&= \mu I_{\lambda^*}(X_S;Y|X_{S^c})+(1-\mu)I_a(X_{S\cup T};Y|X_{S^c\cap T^c}),
\end{align*}
where the term $I_{\lambda^*}(X_S;Y|X_{S^c})$
is calculated with respect to the distribution
\begin{equation*}
  (1-\lambda)p_a(x_{[k]})+\lambda p_b(x_{[k]}).
\end{equation*}
Next, for every $S\subseteq [k]$, define the function
$F_S:[0,h_1)\rightarrow\mathbb{R}$ as
\begin{equation*}
  F_S(h)=I_{\lambda^*}(X_S;Y|U_0,U_{S^c},X_{S^c})-\zeta_{[k]}(h).
\end{equation*}
The following argument shows that for sufficiently small $h$ and for all 
$S,T\subseteq [k]$, 
\begin{equation*}
 f_{S,T}(h)\geq F_{S\cup T}(h).
\end{equation*} 
Consider some $S$ and $T$ for which $T\setminus S$ is not empty.
Then
\begin{align*}
  f_{S,T}(0) &= \mu I_{a}(X_S;Y|X_{S^c})
	+(1-\mu) I_a(X_{S\cup T};Y|X_{S^c\cap T^c})
  +\sum_{j\in T\setminus S}C_\mathrm{in}^j\\
  &\overset{(*)}{>} \mu I_{a}(X_S;Y|X_{S^c})
	+(1-\mu) I_a(X_{S\cup T};Y|X_{S^c\cap T^c})
	+\mu I_{a}(X_{T\setminus S};Y|X_{(T\setminus S)^c})\\
  &\geq I_a(X_{S\cup T};Y|X_{S^c\cap T^c})=F_{S\cup T}(0),
\end{align*}
where $(*)$ follows from (\ref{eq:muDef}). 
Note that $f_{S,T}$ and $F_{S\cup T}$ are continuous functions of $h$ 
for all $S$ and $T$. Thus
there exists $0<h_2\leq h_1$ such that for every $h\in [0,h_2)$
and $S,T\subseteq [k]$ with $T\setminus S\neq\emptyset$,
\begin{equation*}
  f_{S,T}(h)\geq F_{S\cup T}(h).
\end{equation*}
Next consider $S$ and $T$ for which $T\setminus S$ is empty; 
that is, $T$ is a subset of $S$. In this case 
\begin{align*}
  f_{S,T}(h)&=I_{\lambda^*}(X_{S\cup T};Y|U_0,U_{S^c},X_{S^c\cap T^c})
  +\sum_{j\in T\setminus S}C_\mathrm{in}^j-\zeta_{[k]}(h)\\
  &=I_{\lambda^*}(X_S;Y|U_0,U_{S^c},X_{S^c})-\zeta_{[k]}(h)\\
  &=F_S(h)=F_{S\cup T}(h).
\end{align*}
Thus $f_{S,T}(h)\geq F_{S\cup T}(h)$ for all such $S$ and $T$ as well. 
Now fix $h\in [0,h_2)$. From the above argument, it follows that 
the set of all rate vectors that satisfy 
\begin{equation*}
  0\leq\sum_{j\in S}R_j\leq F_S(h)
	\qquad\forall\:\emptyset\neq S\subseteq [k]
\end{equation*}
is achievable. Denote this region
with $\mathscr{C}_\mathrm{ach}(h)$. Now consider the set
of all rate vectors that satisfy 
\begin{equation*}
  0\leq\sum_{j\in S}R_j\leq \Phi_S(h)
	\qquad\forall\:\emptyset\neq S\subseteq [k],
\end{equation*}
where $\Phi_S(h)$ is defined as
\begin{equation*}
  \Phi_S(h)= F_S(h)+\zeta_{S^c}(h)+\sum_{j\in S}C_\mathrm{out}^j(h).
\end{equation*}
Denote this set with $\mathscr{C}_\mathrm{out}(h)$. 
Note that $\mathscr{C}_\mathrm{out}(h)$ is
an outer bound for $\mathscr{C}_\mathrm{ach}(h)$. 

We next show that there exists $0<h_3\leq h_2$ such that for 
every $j\in [k]$ and all $0<h<h_3$, 
\begin{equation} \label{eq:PhijLB}
  \Phi_{\{j\}}(h)>k\sum_{i=1}^k C_\mathrm{out}^i(h).
\end{equation}
To see this, first note that the right hand side of the above
equation equals zero at $h=0$, while
\begin{equation*}
  \Phi_{\{j\}}(0)=I_a(X_j;Y|X_{[k]\setminus\{j\}})>0.
\end{equation*}
Inequality (\ref{eq:PhijLB}) now follows from the fact that
both sides are continuous in $h$. 

By Lemma \ref{lem:submodular}, for a fixed $h$, the mapping $S\mapsto\Phi_S(h)$ is
submodular and nondecreasing. Thus for every $j\in [k]$, there 
exists a rate vector $(R_i)_{i\in [k]}$ in $\mathscr{C}_\mathrm{out}(h)$ such that 
\begin{equation*}
  R_j>k\sum_{i=1}^k C_\mathrm{out}^i(h),
\end{equation*}
and 
\begin{equation*}
  \sum_{j\in [k]}R_j=\Phi_{[k]}(h).
\end{equation*}
For example, for $j=1$, consider the rate vector
$(R_i)_{i\in [k]}$, where $R_1=\Phi_{\{1\}}(h)$, and for all $i>1$,
\begin{equation*}
  R_i=\Phi_{[i]}-\Phi_{[i-1]}.
\end{equation*}
From Corollary 44.3a in \cite[pp. 772]{Schrijver} it follows that 
the defined rate vector is in $\mathscr{C}_\mathrm{out}(h)$.
Now since $\mathscr{C}_\mathrm{out}(h)$ is a convex region, it follows that 
there exists a rate vector $(R_j^*(h))_j$ such that for all $j\in [k]$, 
\begin{equation*}
  R_j^*(h)>\sum_{j=1}^k C_\mathrm{out}^j(h),
\end{equation*}
and 
\begin{equation*}
  \sum_{j=1}^k R_j^*(h)=\Phi_{[k]}(h). 
\end{equation*}

On the other hand, from the definition of $\zeta_S(h)$, given by (\ref{eq:zetaSh}), 
it follows 
\begin{equation*}
  \Phi_S(h)\leq F_S(h)+\sum_{j=1}^k C_\mathrm{out}^j(h).
\end{equation*}
Thus
\begin{equation*}
  \bigg(R_j^*(h)-\sum_{j=1}^k C_\mathrm{out}^j(h)\bigg)_{j\in [k]}
  \in \mathscr{C}_\mathrm{ach}(h).
\end{equation*}
This implies that the sum-rate 
\begin{align*} 
  R_\mathrm{sum}(h) &=\Phi_{[k]}(h)-k\sum_{j=1}^k C_\mathrm{out}^j(h)\\
	&=\mu I_{\lambda^*}(X_{[k]};Y)
	+(1-\mu)I_a(X_{[k]};Y)-k\sum_{j=1}^k C_\mathrm{out}^j(h)
\end{align*}
is achievable. In addition, since 
\begin{equation*}
  R_\mathrm{sum}(0)=I_a(X_{[k]};Y)
  =\max_{p\in\mathcal{P}_\mathrm{ind}(\mathcal{X}_{[k]})}I(X_{[k]};Y),
\end{equation*}
we have 
\begin{equation} \label{eq:Ggf1}
  G(h\mathbf{v})\geq R_\mathrm{sum}(h)-R_\mathrm{sum}(0)
\end{equation}
for all $h\in [0,h_3)$. Thus
\begin{align} \label{eq:GderivLB}
(D_\mathbf{v}G)(\mathbf{0})
&= \lim_{h\rightarrow 0^+}\frac{G(h\mathbf{v})}{h}\notag\\
&\overset{(\mathrm{i})}{\geq} \lim_{h\rightarrow 0^+}
\frac{R_\mathrm{sum}(h)-R_\mathrm{sum}(0)}{h}\notag\\
&= \mu\frac{d}{d{\lambda^*}}I_{\lambda^*}(X_{[k]};Y)
\Big|_{\lambda^*=0^+}\times 
\frac{d\lambda^*}{dh}\Big|_{h=0^+}-k\sum_{j=1}^k v_j\notag\\
&\overset{(\mathrm{ii})}{\geq} 
\frac{\mu}{\epsilon}\bigg[\sum_{x_{[k]}}
\big(p_b(x_{[k]})-p_a(x_{[k]}))D\big(p(y|x_{[k]})\|p_a(y)\big)
\bigg]-k\sum_{j=1}^k v_j.
\end{align}
Here (i) follows from (\ref{eq:Ggf1})
and (ii) is proved by combining (\ref{eq:dlambda})
and Lemma \ref{lem:derivative}, which appears at the
end of this section. From our definitions of $p_a$ 
and $p_b$ it follows
\begin{equation*}
  \sum_{x_{[k]}}p_b(x_{[k]})
	D\big(p(y|x_{[k]})\|p_a(y)\big)
	>\sum_{x_{[k]}}p_a(x_{[k]})
  D\big(p(y|x_{[k]})\|p_a(y)\big).
\end{equation*}
Since $\epsilon$ is arbitrary, from (\ref{eq:GderivLB})
we get
\begin{equation*}
  (D_\mathbf{v}G)(\mathbf{0})=\infty.
\end{equation*}
This completes the proof for the case where 
\begin{equation*}
  S_* :=\big\{j\in [k]:I_a(X_j;Y|X_{[k]\setminus\{j\}})>0\big\}
\end{equation*}
contains $[k]$ (i.e., $S_*=[k]$).
We next consider a MAC for which $S_*$ is a strict subset of $[k]$
(i.e., $S_*\subsetneq [k]$). 

For every $j\in [k]$, let
$\mathcal{A}_j\subseteq\mathcal{X}_j$ denote the the 
support of $p_a(x_j)$. Then for nonempty $S\subseteq [k]$,
the support of $p_a(x_S)$ is given by 
\begin{equation*}
  \mathcal{A}_S=\prod_{j\in S}\mathcal{A}_j.
\end{equation*}
Note that 
\begin{equation*}
  I_a(X_{S_*^c};Y|X_{S_*})
	\leq \sum_{j\in S_*^c}
	I_a(X_j;Y|X_{[k]\setminus\{j\}})=0.
\end{equation*}
Thus for every $x_{S_*}\in\mathcal{A}_{S_*}$,
\begin{equation*}
  I_a(X_{S_*^c};Y|X_{S_*}=x_{S_*})=0,
\end{equation*}
which implies for all $x_{[k]}\in\mathcal{A}_{[k]}$,
\begin{equation*}
  p(y|x_{[k]})=p_a(y|x_{S_*}).
\end{equation*}
Note that since the support of $p_b$ is contained in the 
support of $p_a$ by assumption, it follows that for all
nonempty $S\subseteq [k]$, the support of $p_b(x_S)$ is
contained in $\mathcal{A}_S$. 

Now consider the $|S_*|$-user MAC
\begin{equation*}
  \Big(\mathcal{A}_{S_*},p_a(y|x_{S_*}),\mathcal{Y}\Big),
\end{equation*}
and the input distributions $p_\mathrm{ind}(x_{S_*})=p_a(x_{S_*})$
and $p_\mathrm{dep}(x_{S_*})=p_b(x_{S_*})$.
Note that 
\begin{equation*}
  I_\mathrm{ind}(X_{S_*};Y)
  =\max_{p\in\mathcal{P}(\mathcal{X}_{S_*})}I(X_{S_*};Y),
\end{equation*}
and 
\begin{align*}
  \mathbb{E}_\mathrm{dep}\Big[D\big(p_a(y|X_{S_*})\|p_\mathrm{ind}(y)\big)\Big]
	&= \mathbb{E}_b\Big[D\big(p(y|X_{[k]})\|p_a(y)\big)\Big]\\
	&> \mathbb{E}_a\Big[D\big(p(y|X_{[k]})\|p_a(y)\big)\Big]\\
	&= \mathbb{E}_\mathrm{ind}\Big[D\big(p_a(y|X_{S_*})\|p_\mathrm{ind}(y)\big)\Big].
\end{align*}
Furthermore, for every $j\in S_*$,
\begin{equation*}
  I_\mathrm{ind}(X_j;Y|X_{S_*\setminus\{j\}})=I_a(X_j;Y|X_{[k]\setminus\{j\}})>0.
\end{equation*} 
Thus this MAC satisfies all of the conditions under which we already 
proved Theorem \ref{thm:sumCapacity}. Suppose $\mathbf{v}=(v_j)_{j=1}^k$ is a 
unit vector in $\mathbb{R}^k_{>0}$. Let 
\begin{equation*}
  |\mathbf{v}_{S_*}|=\Big(\sum_{j\in S_*}v_j^2\Big)^{1/2},
\end{equation*}
and define $\mathbf{v}^*=(v^*_j)_{j=1}^k\in\mathbb{R}^k_{>0}$ as
\begin{equation*}
  v^*_j=\frac{v_j}{|\mathbf{v}_{S_*}|}\mathbf{1}\{j\in S_*\}.
\end{equation*}
Then 
\begin{align*}
  (D_\mathbf{v}G)(\mathbf{0}) &= \lim_{h\rightarrow 0^+}\frac{G(h\mathbf{v})}{h}\\
	&\geq |\mathbf{v}_{S_*}|\times \lim_{h\rightarrow 0^+}
	\frac{G(h|\mathbf{v}_{S_*}|\mathbf{v}^*)}{h|\mathbf{v}_{S_*}|}
	\overset{(\star)}{=}\infty, 
\end{align*}
where $(\star)$ follows from the fact that our $|S_*|$-user MAC satisfies all the 
required properties to imply an infinite directional derivative for sum-capacity. 

We next provide
the proofs for the lemmas we use in the above argument. 

The first lemma allows us to simplify the achievable region by replacing the
terms $(R_j-C_\mathrm{in}^j)^+$ with $R_j-C^j_\mathrm{in}$.
\begin{lem} \label{lem:plus}
Let $k$ be a positive integer. Fix $\gamma >0$ and for every $j\in [k]$, 
let $\alpha_j$ be a real 
number. Then the vector $(x_j)_{j\in [k]}$ satisfies
\begin{equation*}
  \sum_{j\in [k]} (x_j-\alpha_j)^+ < \gamma
\end{equation*}
if and only if for every nonempty $S\subseteq [k]$,
\begin{equation*}
  \sum_{j\in S} (x_j-\alpha_j) < \gamma.
\end{equation*}
\end{lem}
\begin{IEEEproof}
Define the sets $\mathcal{A}^+$ and $\mathcal{A}$ as follows
\begin{align*}
  \mathcal{A}&=\Big\{\mathbf{x}\Big|\forall S\subseteq [k]:
  \sum_{j\in S} (x_j-\alpha_j) < \gamma\Big\}\\
  \mathcal{A}^+&=\Big\{\mathbf{x}\Big| 
  \sum_{j\in [k]} (x_j-\alpha_j)^+ < \gamma\Big\}
\end{align*}
Our aim is to show $\mathcal{A}=\mathcal{A}^+$. We first prove
$\mathcal{A}\supseteq\mathcal{A}^+$. 
For every $j\in [k]$, $x_j-\alpha_j\leq (x_j-\alpha_j)^+$; thus
$\mathcal{A}^+\subseteq\mathcal{A}$. We next prove 
$\mathcal{A}\subseteq\mathcal{A}^+$. 
Consider any $\mathbf{x}\in\mathcal{A}$. 
Define the set $S\subseteq [k]$ as
\begin{equation*}
  S=\{j\in [k]|x_j>\alpha_j\}.
\end{equation*}
If $S=\emptyset$, then $\mathbf{x}\in\mathcal{A}^+$ as $\gamma>0$.
If $S$ is not empty, then 
\begin{equation*}
  \sum_{j\in [k]}(x_j-\alpha_j)^+
  =\sum_{j\in S}(x_j-\alpha_j)<\gamma.
\end{equation*}
Thus $\mathbf{x}\in\mathcal{A}^+$. 
\end{IEEEproof}

The next lemma provides the derivative of 
the input-output mutual information and the total 
correlation \cite{Watanabe}, when 
calculated with respect to the convex combination of two 
distributions. In this lemma, $\mathcal{X}_{[k]}$
may be finite, countably infinite, or equal to $\mathbb{R}^k$.
In the first two cases, $p_a$ and $p_b$ are probability mass 
functions. In the case where $\mathcal{X}_{[k]}=\mathbb{R}^k$,
we assume $p_a$ and $p_b$ are ``bounded'' probability density
functions. We say a probability density function $p(x_{[k]})$
on $\mathbb{R}^k$ is bounded if
\begin{equation*}
  \forall\:\emptyset\subsetneq S\subseteq [k]:
	\sup_{\mathcal{X}_S}p(x_S)<\infty. 
\end{equation*}
In addition, in the case where $\mathcal{X}_{[k]}=\mathbb{R}^k$,
the sums should be replaced with integrals. 
\begin{lem} \label{lem:derivative}
Consider two distributions $p_a$ and $p_b$ defined on 
$\mathcal{X}_{[k]}$. For every $\lambda\in [0,1]$, define 
the distribution $p_\lambda$ on $\mathcal{X}_{[k]}$ as
\begin{equation*}
  p_\lambda(x_{[k]})
  =(1-\lambda)p_a(x_{[k]})+\lambda p_b(x_{[k]}). 
\end{equation*}
Then the following statements are true.

(i) For every nonempty $S\subseteq [k]$, we have 
\begin{equation*}
  \frac{d}{d\lambda}H_\lambda(X_S)
  =-\sum_{x_S}(p_b(x_S)-p_a(x_S))\log p_\lambda(x_S).
\end{equation*}

(ii) For every $k$-user MAC 
$(\mathcal{X}_{[k]},p(y|x_{[k]}),\mathcal{Y})$, we have
\begin{equation} \label{eq:dIO}
  \frac{d}{d\lambda}I_\lambda(X_{[k]};Y)
  =\sum_{x_{[k]}}\big(p_b(x_{[k]})-p_a(x_{[k]})\big)
	D\big(p(y|x_{[k]})\|p_\lambda(y)\big).
\end{equation}

(iii) If $p_a$ has the form
\begin{equation*}
  p_a(x_{[k]})=\prod_{j\in [k]}p_a(x_j),
\end{equation*}
and the support of $p_a(x_{[k]})$ contains the support of $p_b(x_{[k]})$,
then for every nonempty $S\subseteq [k]$,
\begin{equation} \label{eq:dTC}
  \frac{d}{d\lambda}\Big(\sum_{j\in S}H_\lambda(X_j)
	-H_\lambda(X_S)\Big)\Big|_{\lambda=0^+}=0.
\end{equation}
\end{lem}

\begin{IEEEproof} 

Claim (i) is clear in the case where $\mathcal{X}_{S}$ is finite. In the
case where $\mathcal{X}_S$ is infinite, we apply the
dominated convergence theorem \cite[p. 55]{Bass}. Define
$f:\mathcal{X}_S\times [0,1]\rightarrow\mathbb{R}$ as 
\begin{equation*}
  f(x_S,\lambda)=p_\lambda(x_S)\log\frac{1}{p_\lambda(x_S)}.
\end{equation*}
Fix $\lambda\in [0,1]$, and consider the sequence of functions $g_n(x_S)$
defined as
\begin{equation*}
  g_n(x_S)=n\big(f(x_S,\lambda+\frac{1}{n})-f(x_S,\lambda)\big).
\end{equation*}
For all $x_S\in\mathcal{X}_S$, we have 
\begin{equation*}
  \lim_{n\rightarrow\infty}g_n(x_S)
	=\frac{\partial f}{\partial \lambda}(x_S,\lambda)
	=-\big(\log e+p_\lambda (x_S)\big)
	\big(p_b(x_S)-p_a(x_S)\big). 
\end{equation*}
By the mean value theorem, for all $x_S\in\mathcal{X}_S$ and $n\in\mathbb{Z}_{>0}$,
there exists $h'\in (0,1/n)$ such that 
\begin{equation*}
  g_n(x_S)=
	\frac{\partial f}{\partial \lambda}(x_S,\lambda+h')=
	-\big(\log e+p_{\lambda+h'}(x_S)\big)
	\big(p_b(x_S)-p_a(x_S)\big). 
\end{equation*}
Since $p_a$ and $p_b$ are bounded, so is $p_{\lambda+h'}$, and
thus, for some constant $C>0$ and all $n\in\mathbb{Z}_{>0}$,
\begin{equation*}
  |g_n(x_S)|\leq C\big|p_b(x_S)-p_a(x_S)\big|.
\end{equation*}
Define $\varphi:\mathcal{X}_S\rightarrow\mathbb{R}$ as 
\begin{equation*}
  \varphi(x_S)=
	C\big|p_b(x_S)-p_a(x_S)\big|.
\end{equation*}
Note that $\varphi\in L^1(\mathcal{X}_S)$, since
\begin{equation*}
  \int_{\mathcal{X}_S}|\varphi(x_S)|dx_S
	\leq 2C. 
\end{equation*}
By the dominated convergence theorem,
\begin{equation*}
  \lim_{n\rightarrow\infty} \sum_{x_S\in\mathcal{X}_S}g_n(x_S)
	=\sum_{x_S\in\mathcal{X}_S}\lim_{n\rightarrow\infty} g_n(x_S),
\end{equation*}
which implies
\begin{equation*}
  \frac{d}{d\lambda}H_\lambda(X_S)
  =-\sum_{x_S}(p_b(x_S)-p_a(x_S))\log p_\lambda(x_S).
\end{equation*}

For (ii), note that
\begin{equation*}
  p_\lambda(y) =(1-\lambda)p_a(y)+\lambda p_b(y).
\end{equation*}
Thus by (i),
\begin{align*}
  \frac{d}{d\lambda}H_\lambda(Y) &=
	-\sum_{y}\big(p_b(y)-p_a(y)\big)(\log e+\log p_\lambda(y))\\
	&= \sum_{y}\big(p_b(y)-p_a(y)\big)\log \frac{1}{p_\lambda(y)}\\ 
	&= \sum_{x_{[k]}}\big(p_b(x_{[k]})-p_a(x_{[k]})\big)
	\sum_y p(y|x_{[k]})\log \frac{1}{p_\lambda(y)}.
\end{align*}
Similarly,
\begin{align*}
  \MoveEqLeft
  \frac{d}{d\lambda}H_\lambda(Y|X_{[k]})\\
	&=\sum_{x_{[k]}}\big(p_b(x_{[k]})-p_a(x_{[k]})\big)
	\sum_y p(y|x_{[k]})\log\frac{1}{p(y|x_{[k]})}.
\end{align*}
Taking the difference between these derivatives 
completes the proof of part (ii). 

For part (iii), note that for every $j\in [k]$, 
\begin{align*}
  \frac{d}{d\lambda}H_\lambda(X_j)
	&=-\sum_{x_j}(p_b(x_j)-p_a(x_j))(\log e+\log p_\lambda(x_j))\\
	&=-\sum_{x_j}(p_b(x_j)-p_a(x_j))\log p_\lambda(x_j)\\
\end{align*}
and
\begin{align*}
  \frac{d}{d\lambda}\sum_{j\in S}H_\lambda(X_j)
	&=-\sum_{j\in S}\sum_{x_j}(p_b(x_j)-p_a(x_j))\log p_\lambda(x_j)\\
	&=\sum_{x_S}(p_b(x_S)-p_a(x_S))\log\frac{1}{\prod_{j\in S}p_\lambda(x_j)}\\
\end{align*}
On the other hand,
\begin{equation*}
  \frac{d}{d\lambda}H_\lambda(X_S)
  =-\sum_{x_S}(p_b(x_S)-p_a(x_S))\log p_\lambda(x_S).
\end{equation*}
Thus
\begin{equation*} 
  \frac{d}{d\lambda}\Big(\sum_{j\in S}H_\lambda(X_j)
	-H_\lambda(X_S)\Big)=
  \sum_{x_S}(p_b(x_S)-p_a(x_S))\log
	\frac{p_\lambda(x_S)}{\prod_{j\in S}p_\lambda(x_j)}.
\end{equation*}
Equation (\ref{eq:dTC}) now follows from the fact that 
\begin{equation*}
  p_a(x_S)=\prod_{j\in S}p_a(x_j),
\end{equation*}
and the support of $p_b$ is contained in the support of 
$p_a$. 
\end{IEEEproof}

In the next lemma, we prove that for a fixed $h$, 
the mapping $S\mapsto\Phi_S(h)$ is nondecreasing 
and submodular. In the statement of this lemma, 
$2^{[k]}$ denotes the collection of all subsets of 
$[k]$. 
\begin{lem} \label{lem:submodular}
Fix a distribution 
\begin{equation*}
 p(u_{[k]})\cdot\prod_{j=1}^k p(x_j|u_j)\cdot p(y|x_{[k]})
\end{equation*}
on $\mathcal{U}_{[k]}\times\mathcal{X}_{[k]}\times\mathcal{Y}$, and
define the function $\Phi:2^{[k]}\rightarrow\mathbb{R}$ as 
\begin{equation*}
 \Phi(S)=I(X_S;Y|U_{S^c}X_{S^c})+
 \sum_{j\in S}H(U_j)-H(U_S|U_{S^c})
\end{equation*}
for every $S\subseteq [k]$.
Then $\Phi$ is nondecreasing and submodular.
\end{lem}
\begin{IEEEproof} 
Note that 
\begin{equation*}
  \Phi(S)=H(Y|U_{S^c}X_{S^c})-H(Y|X_{[k]})
  +\sum_{j\in S}H(U_j)+H(U_{S^c})
  -H(U_{[k]}).
\end{equation*}
For every $j\in {[k]}$, let $V_j=(U_j,X_j)$. Then for 
every $S\subseteq [k]$,
\begin{align*}
 \sum_{j\in S}H(V_j)+H(V_{S^c})-H(V_{[k]})
 &= \sum_{j\in S}H(U_j,X_j)+H(U_{S^c},X_{S^c})-H(U_{[k]},X_{[k]})\\
 &= \sum_{j\in S}H(U_j)+H(U_{S^c})-H(U_{[k]}),
\end{align*}
since each $X_j$ only depends on $U_j$. Thus
\begin{align*}
  \Phi(S) &= H(Y|V_{S^c})-H(Y|V_{[k]})
	+\sum_{j\in S}H(V_j)+H(V_{S^c})-H(V_{[k]})\\
	&=H(V_{S^c}|Y)+\sum_{j\in S}H(V_j)
  -H(V_{[k]}|Y).
\end{align*}

We first show $\Phi$ is nondecreasing. Let $S$ be
a subset of $T$. Then
\begin{align*}
  \MoveEqLeft
  H(V_{S^c}|Y)+\sum_{j\in S}H(V_j)\\
  &= H(V_{T^c}|Y)+H(V_{S^c\setminus T^c}|V_{T^c},Y)
	+\sum_{j\in T}H(V_j)-\sum_{j\in T\setminus S}H(V_j)\\
  &\leq H(V_{T^c}|Y)+\sum_{j\in T}H(V_j),
\end{align*} 
since 
\begin{equation*}
  H(V_{S^c\setminus T^c}|V_{T^c},Y)
	= H(V_{T\setminus S}|V_{T^c},Y)
	\leq \sum_{j\in T\setminus S}H(V_j).
\end{equation*}
Thus $\Phi$ is nondecreasing. 

We next show $\Phi$ is submodular. 
Fix $S,T\subseteq [k]$. Our aim is to prove
\begin{equation} \label{eq:submodular}
  \Phi(S)+\Phi(T)
  \geq\Phi(S\cup T)+\Phi(S\cap T).
\end{equation}
We have
\begin{align*}
  H(V_{S^c}|Y)+H(V_{T^c}|Y)
  &= H(V_{S^c\cap T^c}|Y)+H(V_{S^c\setminus T^c}|V_{S^c\cap T^c},Y)\\
  &\phantom{=}+H(V_{S^c\cup T^c}|Y)-H(V_{S^c\setminus T^c}|V_{T^c},Y)\\
  &= H(V_{S^c\cap T^c}|Y)+H(V_{S^c\cup T^c}|Y)
  +I(V_{S^c\setminus T^c};V_{T^c\setminus S^c}|V_{S^c\cap T^c},Y)\\
  &\geq H(V_{S^c\cap T^c}|Y)+H(V_{S^c\cup T^c}|Y).
\end{align*}
This proves (\ref{eq:submodular}), since
\begin{equation*}
 \sum_{j\in S}H(V_j)+\sum_{j\in T}H(V_j)
 =\sum_{j\in S\cup T}H(V_j)+\sum_{j\in S\cap T}H(V_j).
\end{equation*}
\end{IEEEproof}

\subsection{Proposition \ref{prop:kUserGaussian} (The \texorpdfstring{$k$-}{k}user Gaussian MAC)} 
\label{subsec:kUserGaussian}

For the $k$-user Gaussian MAC, define $p_\mathrm{ind}$ as
\begin{equation*}
  p_\mathrm{ind}(x_{[k]})=\prod_{j\in [k]}\frac{1}{\sqrt{2\pi P_j}}
	\exp\Big(-\frac{x_j^2}{2P_j}\Big)
\end{equation*}
Note that $p_\mathrm{ind}$ satisfies
\begin{equation*}
  I_\mathrm{ind}(X_{[k]};Y)
	=\max_{p\in\mathcal{P}_\mathrm{ind}(\mathcal{X}_{[k]})}
  I(X_{[k]};Y). 
\end{equation*}
From \cite[p. 33]{Pardo},
\begin{align*}
  \MoveEqLeft
	D\big(p(y|x_{[k]})\|p_\mathrm{ind}(y)\big)\\
	&=\frac{1}{2}\left[
	\frac{1}{\sum_{j\in [k]}P_j+N}\Big(\sum_{j\in [k]}x_j\Big)^2
	-\frac{\sum_{j\in [k]}P_j}{\sum_{j\in [k]}P_j+N}+\log\Big(1
	+\frac{1}{N}\sum_{j\in [k]}P_j\Big)\right].
\end{align*}
For $p_\mathrm{dep}$, choose any density function that satisfies
\begin{equation*}
  \forall j\in [k]:
	\mathbb{E}_\mathrm{dep}\big[|X_j|^2\big]\leq P_j
\end{equation*}
and 
\begin{equation} \label{eq:sumAmplitude}
  \mathbb{E}_\mathrm{dep}\bigg[\Big(\sum_{j\in [k]}X_j\Big)^2\bigg]
	>\sum_{j\in [k]}P_j.
\end{equation}
Then (\ref{eq:sumAmplitude}) guarantees 
\begin{align*}
  \mathbb{E}_\mathrm{dep}\Big[D\big(p(y|X_{[k]})\|p_\mathrm{ind}(y)\big)\Big]
	>\mathbb{E}_\mathrm{ind}\Big[D\big(p(y|X_{[k]})\|p_\mathrm{ind}(y)\big)\Big].
\end{align*}
For example, we may choose $p_\mathrm{dep}(x_{[k]})$ to be the distribution
$\mathcal{N}(\mathbf{0},\Sigma)$, where $\Sigma=(\Sigma_{ij})_{i,j\in [k]}$
is given by
\begin{equation*}
  \Sigma_{ij}=
	\begin{cases}
	  \rho\sqrt{P_iP_j} &\text{ if }i\neq j\\
		P_i &\text{ if }i=j,
	\end{cases}
\end{equation*}
where $\rho$ is any number in $(0,1]$. 

\subsection{Proposition \ref{prop:outerBound} (Outer bound)} \label{subsec:outerBoundProof}

Consider a $\big((2^{nR_1},\dots,2^{nR_k}),n,L\big)$-code for the MAC with
a $(\mathbf{C}_\mathrm{in},\mathbf{C}_\mathrm{out})$-CF. For every 
message vector $w_{[k]}=(w_1,\dots,w_k)$, $j\in [k]$, and $\ell\in [L]$, define
\begin{align*}
  u_{j\ell} &= \varphi_{j\ell}(w_j,v_j^{\ell-1})\\
  v_{j\ell} &= \psi_{j\ell}(u_1^\ell,\dots,u_k^\ell),
\end{align*}
where  $u_j^\ell=(u_{j1},\dots,u_{j\ell})$ and 
$v_j^\ell=(v_{j1},\dots,v_{j\ell})$, respectively. 
Also, for every nonempty $S\subseteq [k]$ and $\ell\in [L]$,
let $u_{S\ell}=(u_{j\ell})_{j\in S}$ and $u_{S}^\ell=(u_j^\ell)_{j\in S}$. 
Finally, for every $j\in [k]$, $\ell\in [L]$, and 
$v_j^{\ell-1}\in\mathcal{V}_j^{\ell-1}$, define
the mapping
\begin{align*}
  \varphi^{-1}_{j\ell,v_j^{\ell-1}}:\mathcal{U}_{j\ell}
	&\rightarrow 2^{[2^{nR_j}]}\\
	u_{j\ell} &\mapsto 
	\big\{w_j\big|\varphi_{j\ell}(w_j,v_j^{\ell-1})=u_{j\ell}\big\},
\end{align*}
where $2^{[2^{nR_j}]}$ denotes the set of all the subsets of 
$[2^{nR_j}]$. 

Note that $(v_j^L)_{j=1}^k$ is a deterministic function of
$u_{[k]}^L$. Thus for every $u_{[k]}^L$ and $j\in [k]$, 
the set 
\begin{equation*}
  \mathcal{A}_j(u_{[k]}^L)=\bigcap_{\ell=1}^L
  \varphi_{j\ell,v_j^{\ell-1}}^{-1}(u_{j\ell})
\end{equation*}
is well-defined. It follows that for a fixed code and a given message
vector $w_{[k]}$, the vector of all CF inputs is given by 
$u_{[k]}^L$ if and only if for every $j\in [k]$, $w_j\in\mathcal{A}_j(u_{[k]}^L)$.

By Fano's inequality \cite[p. 38]{CoverThomas}, for some $\epsilon_n=o(1)$, 
\begin{equation*}
  H(W_{[k]}|Y^n)\leq n\epsilon_n.
\end{equation*}
Thus for every nonempty subset $S\subseteq [k]$,
\begin{equation*}
  H(W_S|W_{S^c},U_{[k]}^L,Y^n)\leq n\epsilon_n.
\end{equation*}
We have 
\begin{align} \label{eq:twoMIterms}
  n\sum_{j\in S}R_j &\leq H(W_S|W_{S^c})\notag\\
	&= I(W_S;U_{[k]}^L,Y^n|W_{S^c})+H(W_S|W_{S^c},U_{[k]}^L,Y^n)\notag\\
	&\leq I(W_S;U_{[k]}^L|W_{S^c})+I(W_S;Y^n|W_{S^c},U_{[k]}^L)+n\epsilon_n.
\end{align}
We next find an upper bound for each of the mutual information 
terms. For the first term, we have
\begin{align*}
  I(W_S;U_{[k]}^L|W_{S^c}) &\overset{(a)}{=} H(U_{[k]}^L|W_{S^c})\\
	&=\sum_{\ell=1}^L H(U_{[k]\ell}|W_{S^c},U_{[k]}^{\ell-1})\\
	&=\sum_{\ell=1}^L H(U_{S\ell},U_{S^c\ell}|W_{S^c},U_{[k]}^{\ell-1})\\
	&\overset{(b)}{=}\sum_{\ell=1}^L H(U_{S\ell}|W_{S^c},U_{[k]}^{\ell-1},U_{S^c\ell})
	\leq \sum_{j\in S}C_\mathrm{in}^j,
\end{align*}
where (a) follows from the fact that $U_{[k]}^L$ is a deterministic
function of $W_{[k]}$. Statement (b) follows from the fact that 
$U_{S^c\ell}$ is a deterministic function of $(W_{S^c},U_{[k]}^{\ell-1})$.
For the second term in (\ref{eq:twoMIterms}), we have
\begin{align*}
  I(W_S;Y^n|W_{S^c},U_{[k]}^L) 
	&= H(Y^n|W_{S^c},U_{[k]}^L)-H(Y^n|W_S,W_{S^c},U_{[k]}^L)\\
	&= H(Y^n|U_{[k]}^L,X_{S^c}^n)-H(Y^n|U_{[k]}^L,X_{[k]}^n)\\
	&\leq \sum_{t=1}^n \Big(H(Y_t|X_{S^ct},U_{[k]}^L)-H(Y_t|U_{[k]}^L,X_{[k]t})\Big)\\
	&\leq \sum_{t=1}^n I(X_{St};Y_t|U_{[k]}^L,X_{S^ct}),
\end{align*}
where $X_{St}=(X_{jt})_{j\in S}$. We have 
\begin{equation*}
  p(u_{[k]}^L)=\pr\big\{\forall j\in [k]:W_j\in\mathcal{A}_j(u_{[k]}^L)\big\}
	=\prod_{j=1}^k \frac{|\mathcal{A}_j(u_{[k]}^L)|}{|\mathcal{W}_j|}
\end{equation*}
and
\begin{equation*}
  p(u_{[k]}^L|w_j)=\mathbf{1}\big\{w_j\in\mathcal{A}_j(u_{[k]}^L)\big\}
	\prod_{i\neq j} \frac{|\mathcal{A}_i(u_{[k]}^L)|}{|\mathcal{W}_i|}.
\end{equation*}
Thus
\begin{equation*}
  p(w_j|u_{[k]}^L)=\frac{p(w_j)p(u_{[k]}^L|w_j)}{p(u_{[k]}^L)}
	= \frac{\mathbf{1}\big\{w_j\in\mathcal{A}_j(u_{[k]}^L)\big\}}
	{|\mathcal{A}_j(u_{[k]}^L)|}
\end{equation*}
and
\begin{equation*}
  p(w_{[k]}|u_{[k]}^L)=\frac{p(w_{[k]})p(u_{[k]}^L|w_{[k]})}{p(u_{[k]}^L)}
	= \frac{\prod_{j=1}^k\mathbf{1}\big\{w_j\in\mathcal{A}_j(u_{[k]}^L)\big\}}
	{\prod_{j=1}^k|\mathcal{A}_j(u_{[k]}^L)|}=\prod_{j=1}^k p(w_j|u_{[k]}^L).
\end{equation*}
Therefore, $W_1,\dots,W_k$ are independent given $U_{[k]}^L$. Recall that
at time $t\in [n]$, the output of encoder $j$ is given by $X_{jt}=f_{jt}(W_j,V_j^L)$
for some mapping 
\begin{equation*}
  f_{jt}:[2^{nR_j}]\times\mathcal{V}_j^L
	\rightarrow \mathcal{X}_j.
\end{equation*}
Also define $U_{0t}=U_{[k]}^L$ for all $t\in [n]$. We have
\begin{align*}
  p(x_{[k]t}|u_{0t}) &=\sum_{w_{[k]}}p(w_{[k]}|u_{0t})p(x_{[k]t}|w_{[k]},u_{0t})\\
	&=\sum_{w_{[k]}}\prod_{j=1}^k p(w_{j}|u_{0t})p(x_{jt}|w_{j},u_{0t})\\
	&=\prod_{j=1}^k\sum_{w_{j}} p(w_{j}|u_{0t})p(x_{jt}|w_{j},u_{0t})
	=\prod_{j=1}^k p(x_{jt}|u_{0t}).
\end{align*}
Defining a time sharing random variable and applying the usual time sharing
argument \cite[p. 600]{CoverThomas} completes the proof. 

\subsection{Proposition \ref{prop:gaussianSlope} (The Gaussian MAC)} 
\label{subsec:gaussianSlope}

Consider any $\alpha\in [0,1/2]$. In the region
given in Section \ref{sec:twoUser}, 
set $C_{10}=C_{20}=0$, $C_{1d}=C_{2d}=C_\mathrm{out}$,
$\rho_1=\rho_2=1$, and
\begin{equation*}
  \rho_0=\sqrt{1-2^{-4C_\mathrm{out}}}.
\end{equation*}
Then the rate pair $(R_1^*,R_2^*)$ given by
\begin{align*}
  R_1^* &= \frac{1}{2}\log\Big(
	\frac{1+\gamma_1+\gamma_2+2\rho_0\bar{\gamma}}
	{1+(1-\rho_0^2)\gamma_2}\Big)-C_\mathrm{out}\\
	R_2^* &= \frac{1}{2}\log\big(1+(1-\rho_0^2)\gamma_2\big),
\end{align*}
is achievable. Since
\begin{align*}
  C_\alpha(0) &= \alpha\times\frac{1}{2}
	\log\Big(\frac{1+\gamma_1+\gamma_2}{1+\gamma_2}\Big)
	+(1-\alpha)\times\frac{1}{2}\log(1+\gamma_2)\\
	&=\frac{\alpha}{2}\log(1+\gamma_1+\gamma_2)
	+\frac{1-2\alpha}{2}\log(1+\gamma_2),
\end{align*}
we have
\begin{align} \label{eq:CalphaNearZero}
  \MoveEqLeft 
	C_\alpha(C_\mathrm{out})-C_\alpha(0)\notag\\
	&\geq \alpha R_1^* + (1-\alpha) R_2^* -C_\alpha(0)\\
	&=\frac{\alpha}{2}\log\Big(
	1+\frac{2\rho_0\bar{\gamma}}{1+\gamma_1+\gamma_2}\Big)
	+\frac{1-2\alpha}{2}\log\Big(
	1-\frac{\rho_0^2\gamma_2}{1+\gamma_2}\Big)-C_\mathrm{out}.
\end{align}
Using the fact that $2^x=1+\frac{x}{\log e}+o(x)$ and $\sqrt{1+o(1)}=1+o(1)$, 
we get 
\begin{align*}
  \rho_0 &= \sqrt{1-2^{-4C_\mathrm{out}}}\\
	&=\sqrt{\frac{4C_\mathrm{out}}{\log e}+o(C_\mathrm{out})}\\
	&=\frac{2}{\sqrt{\log e}}\cdot\sqrt{C_\mathrm{out}}
	+o(\sqrt{C_\mathrm{out}}).
\end{align*}
In addition,
\begin{equation*}
  \rho_0^2=\frac{4C_\mathrm{out}}{\log e}+o(C_\mathrm{out})
	=o(\sqrt{C_\mathrm{out}}).
\end{equation*}
Applying $\log(1+x)=x\log e+o(x)$ to (\ref{eq:CalphaNearZero}) completes the proof
for $\alpha\in [0,1/2]$. The proof for $\alpha\in (1/2,1]$ follows similarly. 

\subsection{Proposition \ref{prop:confCF} (Capacity region under the CF and conferencing models)} 
\label{subsec:confCF}

An $L$-round $(C_{ij})_{i,j=1}^k$-conference for a blocklength-$n$ code
is uniquely determined by a collection of sets 
$\{\mathcal{W}_{ij}^{(\ell)}\}_{i,j,\ell}$ and mappings 
\begin{equation*}
 h_{ji}^{(\ell)}:[2^{nR_j}]\times\prod_{i':i'\neq j}\mathcal{W}_{i'j}^{\ell-1}
 \rightarrow \mathcal{W}_{ji}^{(\ell)}
\end{equation*}
where $i,j\in [k]$ and $\ell\in [L]$, and for every $\ell\in [L]$,
\begin{equation*}
 \mathcal{W}_{ij}^\ell
 =\prod_{\ell'=1}^\ell\mathcal{W}_{ij}^{(\ell')}.
\end{equation*}
Furthermore, the sets $W_{ij}^{(\ell)}$ satisfy
\begin{equation*}
\sum_{\ell\in [L]}\log|\mathcal{W}_{ij}^{(\ell)}|
\leq nC_{ij}
\end{equation*}
for all distinct $i,j\in [k]$. Finally, for every message vector
$(m_1,\dots,m_k)$, where $m_j\in [2^{nR_j}]$, define $w_{ji}^{(\ell)}$
recursively as
\begin{equation*}
w_{ji}^{(\ell)}
=h_{ji}^{(\ell)}\Big(m_j,
\big(w_{i'j}^{\ell-1}\big)_{i'\neq j}\Big). 
\end{equation*}
Our aim is to construct a blocklength-$n$ code for the same MAC
with a $(\mathbf{C}_\mathrm{in},\mathbf{C}_\mathrm{out})$-CF that
through $L$ rounds of communication with the encoders, provides
them with the same information as the $L$-round conference given above. 
To this end, for every $j\in [k]$ and $\ell\in [L]$ define the sets
$\mathcal{U}_{j\ell}$ and $\mathcal{V}_{j\ell}$ as
\begin{align*}
\mathcal{U}_{j\ell}
&=\prod_{i:i\neq j}\mathcal{W}_{ji}^{(\ell)}\\
\mathcal{V}_{j\ell}
&=\prod_{i:i\neq j}\mathcal{W}_{ij}^{(\ell)}.
\end{align*}
Then
\begin{align*}
\sum_{\ell=1}^L \log|\mathcal{U}_{j\ell}|
&= \sum_{\ell=1}^L \sum_{i:i\neq j}\log|\mathcal{W}_{ji}^{(\ell)}|\\
&= \sum_{i:i\neq j}\sum_{\ell=1}^L\log|\mathcal{W}_{ji}^{(\ell)}|\\
&\leq n\sum_{i:i\neq j}C_{ji}\leq nC_\mathrm{in}^j.
\end{align*}
Similarly, we show
\begin{equation*}
\sum_{\ell=1}^L \log|\mathcal{V}_{j\ell}|
\leq n\sum_{i:i\neq j}C_{ij}\leq nC_\mathrm{out}^j.
\end{equation*}
Next for every $j\in [k]$ and $\ell\in [L]$, define
the mapping 
\begin{align*}
  \varphi_{j\ell}:[2^{nR_j}]\times\mathcal{V}_j^{\ell-1}
  &\rightarrow\mathcal{U}_{j\ell}\\
  \Big(m_j,\big(w_{ij}^{\ell-1}
  \big)_{i:i\neq j}\Big)
  &\mapsto \big(w_{ji}^{(\ell)}
  \big)_{i:i\neq j}.
\end{align*}
Similarly, define 
\begin{align*}
\psi_{j\ell}:\prod_{i\in [k]}\mathcal{U}_i^{\ell}
&\rightarrow\mathcal{V}_{j\ell}\\
\big(w_{ij'}^{\ell}\big)_{i,j'}
&\mapsto \big(w_{ij}^{(\ell)}
\big)_{i:i\neq j}.
\end{align*}
This completes the proof of the first part.

For the second part, we show that the capacity region of a MAC
with a single-round $(C_{ij})_{i,j}$-conference contains the outer bound
given in Proposition \ref{prop:outerBound} if $C_{ij}\geq C_\mathrm{in}^i$ for all $i,j\in [k]$.
The coding strategy is simple. For each $i\in [k]$, encoder $i$ sends 
the first $nC_\mathrm{in}^i$ bits of its message to all other encoders. The encoders
then form a ``common message,'' that contains the initial $nC_\mathrm{in}^i$ bits of 
message $i$ for all $i\in [k]$. The rest of the proof follows from the forwarding
inner bound (Corollary \ref{cor:forwardInnerBound}) 
with $C_{i0}=C_\mathrm{in}^i$ for all $i\in [k]$. 

\section{Conclusion}

Cooperative strategies allow for a more efficient allocation of network
resources. Here we introduce a model where the encoders of a $k$-user 
MAC cooperate through a larger network. This model allows us to construct
examples of memoryless networks where removing an edge results in a capacity
loss much larger than the capacity of the removed edge, thus proving that
the edge removal property \cite{HoEtAl, JalaliEtAl} 
does not hold for memoryless networks in general.
Finally, we remark that the benefit of cooperation is not limited to achieving higher 
transmission rates, and cooperative strategies also make networks more reliable.
We study the reliability benefit of cooperation in \cite{ZeroCapLong}. 

\appendices

\section{The Multivariate Covering Lemma} \label{app:MCL}

For every positive integer $n$, define the set $[n]=\{1,\dots,n\}$.
Now let $k$ be a positive integer and fix sets
$\mathcal{U}_0,\mathcal{U}_1,\dots,\mathcal{U}_{k+1}$. 
For every nonempty $S\subseteq [k]$ define 
\begin{equation*}
  \mathcal{U}_S=\prod_{j\in S}\mathcal{U}_j.
\end{equation*}
An element of $\mathcal{U}_S$ is denoted with $u_S=(u_j)_{j\in S}$.
Let $p(u_0,u_{[k+1]})$ be a probability distribution on the set 
$\mathcal{U}_0\times\mathcal{U}_{[k+1]}$.
For every $j\in [k]$, let $M_j$ be a nonnegative integer. 
For every nonempty $S\subseteq [k]$, define the set $\mathcal{M}_S$ as
\begin{equation*}
  \mathcal{M}_S=\prod_{j\in S} [M_j].
\end{equation*}
and let $\mathcal{M}=\mathcal{M}_{[k]}$.
For every $\mathbf{m}=(m_1,\dots,m_k)\in \mathcal{M}$,
let the random vector
\begin{equation*}
  (U_0,U_1(m_1),\dots,U_k(m_k),U_{k+1})
\end{equation*} 
have distribution 
\begin{equation*}
  p(u_0)\prod_{j=1}^{k+1} p(u_j|u_0),
\end{equation*} 
where $p(u_0)$ and each $p(u_j|u_0)$ are the conditional 
marginals of $p(u_0,u_{[k+1]})$. 
In addition, let $\mathcal{F}$ be an arbitrary subset of 
$\mathcal{U}_0\times\mathcal{U}_{[k+1]}$. We
want to find upper and lower bounds on the probability
\begin{equation*}
 \pr\Big\{\forall \mathbf{m}\in \mathcal{M}:
	\big(U_0,U_1(m_1),\dots,U_k(m_k),U_{k+1}\big)\notin \mathcal{F}\Big\}.
\end{equation*}
We derive the lower bound (Subsection \ref{subsec:lb}) using the union bound, 
which does not depend on the statistical dependencies of the vectors 
\begin{equation*}
  \big(U_0,U_1(m_1),\dots,U_k(m_k),U_{k+1}\big)
\end{equation*}	
for different values of $\mathbf{m}$. For the upper bound (Subsection \ref{subsec:ub}), 
which leads to the multivariate covering lemma, we require a stronger assumption, 
which we next describe.
  
Let $\mathbf{m}$ and $\mathbf{m}'$ be in $\mathcal{M}$. Define the set
$S_{\mathbf{m},\mathbf{m}'}$ as
\begin{equation*}
  S_{\mathbf{m},\mathbf{m}'}=\big\{j\in [k]:m_j=m'_j\big\}.
\end{equation*}
When $\mathbf{m}$ and $\mathbf{m}'$ are clear from context, we denote 
$S_{\mathbf{m},\mathbf{m}'}$ with $S$. In the proof of the upper bound
we require 
\begin{align*}
  \MoveEqLeft
  \pr\Big\{\forall j\in [k]:U_j(m_j)=u_j\text{ and }U_j(m'_j)=u'_j\Big|
  U_0=u_0,U_{k+1}=u_{k+1}\Big\}\\
  &= \prod_{j=1}^k p(u_j|u_0)\times \prod_{j\in S^c}p(u'_j|u_0),
\end{align*}
for all $u_0$ and
all $(u_j)_j$ and $(u'_j)_j$ such that if $j\in S$, then $u_j=u'_j$ (Assumption I). Note 
that if there exists a $j\in S$ where $u_j\neq u'_j$ then the probability on the 
left hand side equals zero.

In the corresponding asymptotic problem (Subsection \ref{subsec:asymp}), 
we apply our bounds to 
\begin{equation*}
  \pr\Big\{\forall \mathbf{m}:
	\big(U_0^n,U_1^n(m_1),\dots,U_k^n(m_k),
	U_{k+1}^n\big)\notin A_\delta^{(n)}\Big\},
\end{equation*} 
where for every $\mathbf{m}$,
\begin{equation*}
  \big(U_0^n,U_1^n(m_1),\dots,U_k^n(m_k),
  U_{k+1}^n\big)
\end{equation*}	
is simply $n$ i.i.d.\ copies of the original random vector
\begin{equation*}
  \big(U_0,U_1(m_1),\dots,U_k(m_k),
  U_{k+1}\big),
\end{equation*}	
(Assumption II) and $A_\delta^{(n)}$ is the weakly typical set \cite[p. 521]{CoverThomas} 
defined with respect to the distribution $p(u_0,u_{[k+1]})$. 
The multivariate covering lemma follows. 

\begin{lem}[Multivariate Covering Lemma] \label{lem:MCL}
Suppose assumptions (I) and (II) hold for the joint distribution of
\begin{equation*}
  U_0^n,\big\{U_1^n(m_1),\dots,U_k^n(m_k)\big\}_\mathbf{m},U_{k+1}^n.
\end{equation*}
For the direct part, suppose for all $j\in [k]$, $M_j\geq 2^{nR_j}$. If  
for all nonempty $S\subseteq [k]$,
\begin{equation} \label{eq:sumrj}
  \sum_{j\in S} R_j >
  \sum_{j\in S}H(U_j|U_0)-H(U_S|U_0,U_{k+1})+(8k-2|S|+10)\delta,
\end{equation}
then
\begin{equation} \label{eq:limit}
  \lim_{n\rightarrow \infty}
	\pr\Big\{\exists \mathbf{m}:\big(U_0^n,U_1^n(m_1),\dots,U_k^n(m_k),U_{k+1}^n\big)
	\in A_\delta^{(n)}\Big\}=1.
\end{equation}
For the converse, assume for all $j\in [k]$, $M_j\leq 2^{nR_j}$. 
If (\ref{eq:limit}) holds, then
\begin{equation*}
  \sum_{j\in S} R_j \geq 
  \sum_{j\in S}H(U_j|U_0)-H(U_S|U_0,U_{k+1})-2(|S|+1)\delta,
\end{equation*}
for all nonempty $S\subseteq [k]$. 
\end{lem}

\textbf{Remark.} In the direct part of Lemma \ref{lem:MCL}, we can weaken the lower bound
on $\sum_{j\in S} R_j$ when $S=[k]$. Specifically, 
we can replace (\ref{eq:sumrj}) with 
\begin{equation*}
  \sum_{j=1}^k R_j > \sum_{j=1}^k H(U_j|U_0)-H(U_{[k]}|U_0,U_{k+1})+2(k+1)\delta.
\end{equation*}
for $S=[k]$. 

\subsection{The Lower Bound} \label{subsec:lb}

Define the distribution $p_\mathrm{ind}(u_0,u_{[k+1]})$ on the
set $\mathcal{U}_0\times\mathcal{U}_{[k+1]}$ as 
\begin{equation*}
  p_\mathrm{ind}(u_0,u_{[k+1]})
	=p(u_0,u_{k+1})\prod_{j\in [k]}p(u_j|u_0). 
\end{equation*}
For every $S\subseteq [k]$,
define $\mathcal{F}_S$ as the projection of $\mathcal{F}$ on
$\mathcal{U}_0\times\mathcal{U}_S\times\mathcal{U}_{k+1}$, and
for every $(u_0,u_S,u_{k+1})\in\mathcal{F}_S$, 
let $\mathcal{F}(u_0,u_S,u_{k+1})$ be the set of all $u_{S^c}$ 
such that $(u_0,u_{[k+1]})\in\mathcal{F}$.
In addition, for every nonempty $S\subseteq [k]$, let $\alpha_S$ and $\beta_S$ 
be constants such that for all $(u_0,u_S,u_{k+1})\in \mathcal{F}_S$
\begin{equation*}
  \alpha_S \leq 
  \log \frac{p(u_S|u_0,u_{k+1})}{p_\mathrm{ind}(u_S|u_0)},
\end{equation*}
and for all $(u_0,u_S,u_{S^c},u_{k+1})\in \mathcal{F}$, 
\begin{equation*}
  \beta_S \leq \log 
	\frac{p(u_S|u_0,u_{S^c},u_{k+1})}{p_\mathrm{ind}(u_S|u_0)}.
\end{equation*}
Furthermore, let the constant $\gamma$ satisfy
\begin{equation*}
  \gamma \geq \log \frac{p(u_{[k]}|u_0,u_{k+1})}
	{p_\mathrm{ind}(u_{[k]}|u_0)}
\end{equation*}
for all $(u_0,u_{[k]},u_{k+1})\in\mathcal{F}$. 

For every $\mathbf{m}=(m_1,\dots,m_k)\in \mathcal{M}$, define the 
random variable $Z_\mathbf{m}$ as
\begin{equation*}
  Z_\mathbf{m}=\mathbf{1}\Big\{\big(U_0,U_1(m_1),\dots,U_k(m_k),
  U_{k+1}\big)\in \mathcal{F}\Big\}
\end{equation*}
and set 
\begin{equation*}
  Z=\sum_{\mathbf{m}\in\mathcal{M}}Z_\mathbf{m}.
\end{equation*}
Our aim is to find a lower bound for $\pr\{Z=0\}$. 
Note that for every nonempty $S\subseteq [k]$, 
\begin{align*}
  \pr\big\{\exists \mathbf{m}:Z_\mathbf{m}=1\big\}
	&= \pr\Big\{\exists \mathbf{m}: \big(U_0,U_1(m_1),\dots,U_k(m_k),U_{k+1}\big)\in \mathcal{F}\Big\}\\
	&\leq \pr\Big\{\exists \mathbf{m}: \big(U_0,\big(U_j(m_j)\big)_{j\in S},U_{k+1}\big)\in\mathcal{F}_S\Big\}\\
	&\leq |\mathcal{M}_S|\sum_{\mathcal{F}_S}p(u_0,u_{k+1})p_\mathrm{ind}(u_S|u_0)\\
	&\leq |\mathcal{M}_S|2^{-\alpha_S}\sum_{\mathcal{F}_S}p(u_0,u_S,u_{k+1})\\
	&\leq |\mathcal{M}_S|2^{-\alpha_S}.
\end{align*}
Thus 
\begin{align} \label{eq:lb}
  \pr\{Z=0\} &= 1-\pr\big\{\exists \mathbf{m}:Z_\mathbf{m}=1\big\}\notag\\
	&\geq 1-\min_{|S|\neq\emptyset}|\mathcal{M}_S|2^{-\alpha_S}.
\end{align}

\subsection{The Upper Bound} \label{subsec:ub}

In deriving our upper bound on $\pr\{Z=0\}$, we apply conditioning
and Chebyshev's inequality. Thus, the factor 
\begin{equation*}
  \frac{1}{\big(\pr\{\mathcal{F}(u_0,u_{k+1})\}\big)^2}
\end{equation*}
appears, where 
\begin{align*}
  \pr\big\{\mathcal{F}(u_0,u_{k+1})\big\} &=
  \pr\Big\{U_{[k]}\in \mathcal{F}(u_0,u_{k+1})\Big|U_0=u_0,U_{k+1}=u_{k+1}\Big\}\\
  &=\sum_{u_{[k]}\in\mathcal{F}(u_0,u_{k+1})}p(u_{[k]}|u_0,u_{k+1})
\end{align*} 
and $\mathcal{F}(u_0,u_{k+1})$ (Subsection \ref{subsec:lb}) is simply the set of all
$u_{[k]}\in\mathcal{U}_{[k]}$ that satisfy $(u_0,u_{[k]},u_{k+1})\in\mathcal{F}$. 
Thus to get a reasonably accurate upper bound, we require
$\pr\{\mathcal{F}(u_0,u_{k+1})\}$ to be large. However, as we cannot guarantee this
for all $(u_0,u_{k+1})$, we partition the $(u_0,u_{k+1})$ pairs into ``good'' and ``bad'' sets, 
corresponding to large and small values of 
$\pr\{\mathcal{F}(u_0,u_{k+1})\}$, respectively.
The probability of the good set is large when
$\pr\{(U_0,U_{[k+1]})\in\mathcal{F}\}$ 
is sufficiently large. To see this,
fix $\epsilon>0$. Following Appendix III of \cite{KoetterEtAl2}, define the
set $\mathcal{G}\subseteq\mathcal{U}_0\times\mathcal{U}_{k+1}$ as
\begin{equation*}
  \mathcal{G}=\big\{(u_0,u_{k+1}):\pr\{\mathcal{F}(u_0,u_{k+1})\}\geq 1-\epsilon\big\},
\end{equation*}
Note that $\mathcal{G}$ is the set of all good $(u_0,u_{k+1})$ pairs as defined
above. We have
\begin{align*}
  \pr\big\{(U_0,U_{[k+1]})\in\mathcal{F}\big\} 
  &= \sum_{u_0,u_{k+1}}p(u_0,u_{k+1})\pr\{\mathcal{F}(u_0,u_{k+1})\}\\
  &\leq (1-\epsilon)\pr\{(U_0,U_{k+1})\notin\mathcal{G}\}+\pr\{(U_0,U_{k+1})\in\mathcal{G}\}\\
  &= 1-\epsilon \pr\{(U_0,U_{k+1})\notin\mathcal{G}\}.
\end{align*}
Thus
\begin{equation} \label{eq:notinV}
  \pr\{(U_0,U_{k+1})\notin\mathcal{G}\}\leq 
  \frac{1}{\epsilon}\pr\big\{(U_0,U_{[k+1]})\notin\mathcal{F}\big\}.
\end{equation}
Our aim is to find an upper bound for $\pr\{Z=0\}$. To do this, we write
\begin{align} \label{eq:initialub}
  \pr\{Z=0\} &= \sum_{u_0,u_{k+1}}p(u_0,u_{k+1})\pr\{Z=0|u_0,u_{k+1}\} \notag\\
  &\leq \frac{1}{\epsilon}\pr\big\{(U_0,U_{[k]},U_{k+1})\notin\mathcal{F}\big\}+
  \sum_{(u_0,u_{k+1})\in \mathcal{G}}p(u_0,u_{k+1})\pr\{Z=0|u_0,u_{k+1}\},
\end{align}
where the inequality follows from (\ref{eq:notinV}).
Therefore, to find an upper bound on $\pr\{Z=0\}$, it suffices
to find an upper bound on $\pr\{Z=0|u_0,u_{k+1}\}$
for all $(u_0,u_{k+1})\in \mathcal{G}$. 

Fix $(u_0,u_{k+1})\in \mathcal{G}$.
We use Chebyshev's inequality to find an upper bound on 
$\pr\{Z=0|u_0,u_{k+1}\}$. Thus we 
need to calculate $\mathbb{E}[Z|u_0,u_{k+1}]$ 
and $\mathbb{E}[Z^2|u_0,u_{k+1}]$. For a given $\mathbf{m}$,
from the definition of $\gamma$ (Subsection \ref{subsec:lb}) it follows
\begin{align*}
  \mathbb{E}[Z_\mathbf{m}|u_0,u_{k+1}]
  &=\pr\Big\{\big(U_1(m_1),\dots,U_k(m_k)\big)\in 
  \mathcal{F}(u_0,u_{k+1})\big|u_0,u_{k+1}\Big\}\\
  &=\sum_{\mathcal{F}(u_0,u_{k+1})}p_\mathrm{ind}(u_{[k]}|u_0)\\
  &\geq \sum_{\mathcal{F}(u_0,u_{k+1})}2^{-\gamma}
  p(u_{[k]}|u_0,u_{k+1})\\
  &= 2^{-\gamma}\pr\{\mathcal{F}(u_0,u_{k+1})\}\geq (1-\epsilon)2^{-\gamma}.
\end{align*}
where the last inequality follows from the fact that $(u_0,u_{k+1})\in\mathcal{G}$.
Thus, by linearity of expectation,
\begin{equation} \label{eq:explb}
  \mathbb{E}[Z|u_0,u_{k+1}]\geq |\mathcal{M}|2^{-\gamma}(1-\epsilon).
\end{equation}
Next, we find an upper bound on $\mathbb{E}[Z^2|u_0,u_{k+1}]$. We have
\begin{equation*}
  Z^2 = \sum_{\mathbf{m}}Z_\mathbf{m}^2+
  \sum_{\mathbf{m}\neq\mathbf{m}'} Z_{\mathbf{m}}Z_{\mathbf{m}'}
	=Z+\sum_{\mathbf{m}\neq\mathbf{m}'} Z_{\mathbf{m}}Z_{\mathbf{m}'},
\end{equation*}
since $Z_\mathbf{m}^2=Z_\mathbf{m}$ and $Z=\sum_{\mathbf{m}}Z_\mathbf{m}$.
Thus 
\begin{equation*}
  \mathbb{E}[Z^2|u_0,u_{k+1}]=\mathbb{E}[Z|u_0,u_{k+1}]+\mathbb{E}
	\Big[\sum_{\mathbf{m}\neq\mathbf{m}'} Z_{\mathbf{m}}Z_{\mathbf{m}'}\Big|u_0,u_{k+1}\Big]
\end{equation*}
For any pair of distinct $\mathbf{m}$ and $\mathbf{m}'$ with 
nonempty $S=S_{\mathbf{m},\mathbf{m}'}$, we have
\begin{align*}
  \MoveEqLeft
  \mathbb{E}\big[Z_{\mathbf{m}}Z_{\mathbf{m}'}|u_0,u_{k+1}\big]\\
  &=\sum_{\mathcal{F}_S(u_0,u_{k+1})}p_\mathrm{ind}(u_S|u_0)\bigg[
  \sum_{u_{S^c}\in \mathcal{F}(u_0,u_S,u_{k+1})}p_\mathrm{ind}(u_{S^c}|u_0)\bigg]^2\\
  &\leq 2^{-\alpha_S-2\beta_{S^c}}
  \sum_{\mathcal{F}_S(u_0,u_{k+1})}p(u_S|u_0,u_{k+1})\bigg[
  \sum_{u_{S^c}\in \mathcal{F}(u_0,u_S,u_{k+1})}p(u_{S^c}|u_0,u_S,u_{k+1})\bigg]^2\\
  &\leq 2^{-\alpha_S-2\beta_{S^c}},
\end{align*}
where $\mathcal{F}_S(u_0,u_{k+1})$ is the set of all $u_S$ that satisfy 
$(u_0,u_S,u_{k+1})\in \mathcal{F}_S$.
On the other hand, if $S=S_{\mathbf{m},\mathbf{m}'}$ is empty, then
$Z_\mathbf{m}$ and $Z_\mathbf{m}'$ are independent given 
$(U_0,U_{k+1})=(u_0,u_{k+1})$, and  
\begin{equation*}
  \mathbb{E}\big[Z_{\mathbf{m}}Z_{\mathbf{m}'}|u_0,u_{k+1}\big]
	= \big(\mathbb{E}[Z_\mathbf{m}|u_0,u_{k+1}]\big)^2.
\end{equation*}
Thus
\begin{align} \label{eq:varub}
  \mathbb{E}[Z^2|u_0,u_{k+1}] &= \mathbb{E}[Z|u_0,u_{k+1}]
  +\big(\mathbb{E}[Z|u_0,u_{k+1}]\big)^2 \notag\\
  &\phantom{=} +\sum_{\emptyset\subset S\subset [k]} |\mathcal{M}_S| \prod_{j\in S^c}
  \big(|\mathcal{M}_j|^2-|\mathcal{M}_j|\big)\mathbb{E}[Z_{\mathbf{m}}Z_{\mathbf{m}'}|u_0,u_{k+1}]\notag\\
  &\leq \mathbb{E}[Z|u_0,u_{k+1}]+\big(\mathbb{E}[Z|u_0,u_{k+1}]\big)^2+
  \sum_{\emptyset\subset S\subset [k]} 
  |\mathcal{M}_S||\mathcal{M}_{S^c}|^2 2^{-\alpha_S-2\beta_{S^c}},
\end{align}
where the notation $\emptyset\subset S\subset [k]$ means that $S$ is a nonempty proper
subset of $[k]$. Thus for all $(u_0,u_{k+1})\in\mathcal{G}$, we have
\begin{align*}
 \pr\big\{Z=0|u_0,u_{k+1}\big\}&\leq 
 \pr\Big\{\big|Z-\mathbb{E}[Z|u_0,u_{k+1}]\big|
 \geq \mathbb{E}[Z|u_0,u_{k+1}]\Big|u_0,u_{k+1}\Big\}\\
 &\overset{(a)}{\leq} \frac{\mathrm{Var}(Z|u_0,u_{k+1})}{\big(\mathbb{E}[Z|u_0,u_{k+1}]\big)^2}=
 \frac{\mathbb{E}[Z^2|u_0,u_{k+1}]}{\big(\mathbb{E}[Z|u_0,u_{k+1}]\big)^2}-1\\
 &\overset{(b)}{\leq} \frac{1}{1-\epsilon} |\mathcal{M}|^{-1}2^\gamma
 +\frac{1}{(1-\epsilon)^2}\sum_{\emptyset\subset S\subset [k]}|\mathcal{M}_S|^{-1}
 2^{-\alpha_S-2\beta_{S^c}+2\gamma},
\end{align*}
where (a) follows from Chebyshev's inequality and
(b) follows from (\ref{eq:explb}) and (\ref{eq:varub}). 
Now using (\ref{eq:initialub}), we get
\begin{equation} \label{eq:ub}
  \pr\{Z=0\}\leq \frac{1}{\epsilon}\pr\{\mathcal{F}^c\}+
	\frac{1}{1-\epsilon}|\mathcal{M}|^{-1}2^\gamma
  +\frac{1}{(1-\epsilon)^2}
 \sum_{\emptyset\subset S\subset [k]}|\mathcal{M}_S|^{-1}
 2^{-\alpha_S-2\beta_{S^c}+2\gamma}.
\end{equation}

\subsection{The Asymptotic Result} \label{subsec:asymp}

In this section, using our lower and upper bounds, we prove Lemma \ref{lem:MCL}.
We first prove the direct part using our upper 
bound from Section \ref{subsec:ub}. 
Set $\mathcal{F}=A_\delta^{(n)}$ and for every $j\in [k]$,
choose an integer $M_j\geq 2^{nR_j}$. 
Choose a sequence $\{\epsilon_n\}_n$ such
that
\begin{equation*}
  \lim_{n\rightarrow\infty}
  \frac{1}{\epsilon_n}\pr\big\{(A_\delta^{(n)})^c\big\}=0.
\end{equation*}
Fix a nonempty $S\subseteq [k]$. 
Notice that if
$\big(U_0^n,(U_j^n)_{j\in S},U_{k+1}^n\big)\in \mathcal{F}_S$, then
\begin{equation*}
  \Big|\log\frac{p(u_S^n|u_0^n,u_{k+1}^n)}{\prod_{j\in S}p(u_j^n|u_0^n)}
	-n\Big(\sum_{j\in S}H(U_j|U_0)-H(U_S|U_0,U_{k+1})\Big)\Big|
	\leq 2n(|S|+1)\delta.
\end{equation*}
Thus we may choose 
\begin{equation*}
  \alpha_S = n\Big(\sum_{j\in S}H(U_j|U_0)-H(U_S|U_0,U_{k+1})-2(|S|+1)\delta\Big)
\end{equation*}
and 
\begin{equation*}
  \gamma = n\Big(\sum_{j=1}^k H(U_j|U_0)-H(U_{[k]}|U_0,U_{k+1})+2(k+1)\delta\Big).
\end{equation*}
Similarly, for every nonempty $S\subseteq [k]$, we choose 
$\beta_{S}$ as
\begin{equation*}
  \beta_{S} = n\Big(\sum_{j\in S}H(U_j|U_0)-H(U_S|U_0,U_{S^c},U_{k+1})
  -2(|S|+1)\delta)\Big),
\end{equation*}
since for every $\big(U_0^n,(U_j^n)_{j\in S},(U_j^n)_{j\in S^c}\big)\in \mathcal{F}$,
\begin{equation*}
  \Big|\log\frac{p(u_S^n|u_0^n,u_{S^c}^n,u_{k+1}^n)}{\prod_{j\in S}p(u_j^n|u_0^n)}
	-n\Big(\sum_{j\in S}H(U_j|U_0)-H(U_S|U_0,U_{S^c},U_{k+1})\Big)\Big|
	\leq 2n(|S|+1)\delta.
\end{equation*}
From our upper bound, Equation (\ref{eq:ub}), it now follows that if 
for all nonempty $S\subset [k]$,
\begin{align*}
  \sum_{j\in S} R_j &> \frac{1}{n}(2\gamma-\alpha_S-2\beta_{S^c})\\
	&= 2\sum_{j=1}^k H(U_j|U_0)-2H(U_{[k]}|U_0,U_{k+1})
	-\sum_{j\in S}H(U_j|U_0)+H(U_S|U_0,U_{k+1})\\
	&\phantom{=}-2\sum_{j\in S^c}H(U_j|U_0)+2H(U_{S^c}|U_0,U_S,U_{k+1})
  +(8k-2|S|+10)\delta\\
	&=\sum_{j\in S}H(U_j|U_0)-H(U_S|U_0,U_{k+1})+(8k-2|S|+10)\delta,
\end{align*}
and for $S=[k]$,
\begin{equation*}
  \sum_{j=1}^k R_j > \frac{1}{n}\gamma=\sum_{j=1}^k H(U_j|U_0)
  -H(U_{[k]}|U_0,U_{k+1})-2(k+1)\delta,
\end{equation*}
then  
\begin{equation} \label{eq:limit4}
  \lim_{n\rightarrow \infty}
	\pr\Big\{\exists \mathbf{m}:\big(U_0^n,U_1^n(m_1),\dots,U_k^n(m_k),U_{k+1}^n\big)
	\in A_\delta^{(n)}\Big\}=1.
\end{equation}

Next we prove the converse. Suppose for each $j\in [k]$,
$M_j\leq 2^{nR_j}$ and (\ref{eq:limit4}) holds. Then
from our lower bound, Equation (\ref{eq:lb}), it follows
\begin{equation*}
  \sum_{j\in S} R_j \geq \frac{1}{n}\alpha_S
	= \sum_{j\in S}H(U_j|U_0)-H(U_S|U_0,U_{k+1})-2(|S|+1)\delta,
\end{equation*}
for all nonempty $S\subseteq [k]$. 

\section{Large Deviations} \label{app:largeDev}

In this appendix, we state and prove the following result. It is 
well known and is included for completeness. 
\begin{lem} \label{lem:largeDev}
Choose a distribution $p(u_{[k]})$ on the alphabet $\mathcal{U}_{[k]}$,
which may be continuous or discrete. Suppose there exists $t_0>0$ so
that for all nonempty $S\subseteq [k]$ and $t\in (-t_0,t_0)$,
\begin{equation*}
  \mathbb{E}\big[p(U_S)^{-t}\big]<\infty.
\end{equation*}
Then there exists a nondecreasing function 
$I:\mathbb{R}_{>0}\rightarrow\mathbb{R}_{>0}$ such that for all 
sufficiently large $n$, 
\begin{equation*}
  \pr\big\{A_\epsilon^{(n)}(U_{[k]})\big\}
	\geq 1-2^{-nI(\epsilon)}.
\end{equation*}
\end{lem}
\begin{IEEEproof}
The moment generating function of a random variable $X$ is defined as
\begin{equation*}
  M(t)=\mathbb{E}[e^{tX}]
\end{equation*}
for all real $t$ for which the expectation on the right hand side
is finite. If $M$ is defined on a neighborhood of $0$, say 
$(-t_1,t_1)$ for some $t_1>0$, then it has a Taylor series expansion
with a positive radius of convergence \cite[pp. 278-280]{Billingsley}. In 
particular, 
\begin{equation*}
  \frac{d}{dt}M(t)\big|_{t=0}=\mathbb{E}[X].
\end{equation*}

We next find an upper bound for $\pr\{X\geq a\}$ for
any $a\in\mathbb{R}$. Choose $t\in (0,t_1)$. Using Markov's inequality,
we get
\begin{align*}
  \pr\{X\geq a\} &= \pr\{tX\geq ta\}\\
	&=\pr\{e^{tX}\geq e^{ta}\}\\
	&\leq e^{-ta}\mathbb{E}[e^{tX}]\\
	&= e^{\log M(t)-ta}
\end{align*}
Since $t\in (0,t_1)$ was arbitrary, we get
\begin{equation*}
  \pr\{X\geq a\}\leq e^{\inf_{t\in (0,t_1)}(\ln M(t)-ta)}.
\end{equation*} 
Define the function $f$ as
\begin{equation*}
  f(t)=\ln M(t)-ta.
\end{equation*}
Then $f(0)=0$ and $f'(0)=\mathbb{E}[X]-a$. Thus if
$a>\mathbb{E}[X]$, 
\begin{equation} \label{eq:legendre}
  \inf_{t\in (0,t_1)}\big(\ln M(t)-ta\big)<0.
\end{equation}
If we apply the same inequality to the random variable
\begin{equation*}
  \frac{1}{n}\sum_{i=1}^n X_i,
\end{equation*}
where the $X_i$'s are i.i.d.\ copies of $X$, we get 
\begin{equation} \label{eq:Chernoff}
  \pr\Big\{\sum_{i=1}^n X_i\geq na\Big\}\leq e^{n\inf_{t\in (0,t_1)}(\ln M(t)-ta)}.
\end{equation} 

Now consider a random vector $(U_1,\dots,U_k)$ with distribution 
$p(u_1,\dots,u_k)$. For every nonempty $S\subseteq [k]$, let
$U_S$ denote the random vector $(U_j)_{j\in S}$. 
Let $(U_1^n,\dots,U_k^n)$ be $n$ i.i.d.\ copies of $(U_1,\dots,U_k)$. By applying
inequality (\ref{eq:Chernoff}) to the random variables 
$\{\log\frac{1}{p(U_{Si})}\}_{i=1}^n$ and setting $a=H(U_S)+\epsilon$ for 
some $\epsilon>0$, we get
\begin{equation}
  \pr\Bigg\{\sum_{i=1}^n \log\frac{1}{p(U_{Si})}\geq n(H(U_S)+\epsilon)\Bigg\}
	\leq 2^{-nI_S(\epsilon)},
\end{equation} 
where $I_S(\epsilon)$ is given by
\begin{equation} \label{eq:ISeps}
  I_S(\epsilon)=\inf_{t\in (0,t_0)}
	\Big\{\frac{t}{\ln 2}\big(H(U_S)+\epsilon\big)-\log\mathbb{E}\big[p(U_S)^{-t}\big]\Big\}
\end{equation}
Let
\begin{equation*}
  I(\epsilon)=\frac{1}{2}\min_{S\subseteq [k]}I_S(\epsilon).
\end{equation*}
By the union bound we get 
\begin{align*}
  \pr\big\{(U_1^n,\dots,U_k^n)\notin 
	A_\epsilon^{(n)}(U_1,\dots,U_k)\big\}
	&\leq 2\sum_{\emptyset\subsetneq S\subseteq [k]}e^{-nI_S(\epsilon)}\\
	&\leq 2(2^k-1)2^{-n\min_S I_S(\epsilon)}\\
	&\leq 2^{-nI(\epsilon)},
\end{align*}
where the last inequality holds for all sufficiently large $n$. 
Finally, note that since by (\ref{eq:legendre}) and (\ref{eq:ISeps}), 
each $I_S(\epsilon)$ is positive and nondecreasing, so is $I(\epsilon)$.
\end{IEEEproof}

\section*{Acknowledgment}
The first author thanks M. F. Wong for useful discussions 
regarding the proof of Theorem \ref{thm:sumCapacity}.

\bibliographystyle{IEEEtran}
\bibliography{ref}{}

\end{document}